\documentclass[twocolumn]{aastex63}
\usepackage{natbib}
\newcommand\mname{DYNAMITE}

\received{}
\revised{}
\accepted{}
\submitjournal{AJ}

\shorttitle{Hidden Worlds}
\shortauthors{Dietrich \& Apai}
\graphicspath{{./}{}}
\begin{document}

\defcitealias{mul18}{M18}
\defcitealias{hef19}{H19}

%\title{Hidden Worlds: A Method for Dynamically Predicting Undetected Planets in Multi-planet Systems and Its Applications to \textit{TESS} Systems}

\title{Hidden Worlds: Dynamical Architecture Predictions of Undetected Planets in Multi-planet Systems and Applications to \textit{TESS} Systems}

\correspondingauthor{Jeremy Dietrich}
\email{jdietrich1@email.arizona.edu}

\author[0000-0001-6320-7410]{Jeremy Dietrich}
\affiliation{Department of Astronomy, The University of Arizona, Tucson, AZ 85721, USA}

\author[0000-0003-3714-5855]{D\'aniel Apai}
\affiliation{Department of Astronomy, The University of Arizona, Tucson, AZ 85721, USA}
\affiliation{Lunar and Planetary Laboratory, The University of Arizona, Tucson, AZ 85721, USA}
%\collaboration{1}{(Project EDEN)}

%% depreciated in this version as it is no longer necessary.  AASTeX 

%% provide the collaboration status of a group of authors.  These commands 
%% can be used either before or after the list of corresponding authors.  The
%% argument for \collaboration is the collaboration identifier.  Authors are
%% encouraged to surround collaboration identifiers with ()s.  The 

\begin{abstract}

Multi-planet systems produce a wealth of information for exoplanet science, but our understanding of planetary architectures is incomplete.  Probing these systems further will provide insight into orbital architectures and formation pathways.  Here we present a model to predict previously undetected planets in these systems via population statistics.  The model considers both transiting and non-transiting planets, and can test the addition of more than one planet.  Our tests show the model's orbital period predictions are robust to perturbations in system architectures on the order of a few percent, much larger than current uncertainties.  Applying it to the multi-planet systems from \textit{TESS} provides a prioritized list of targets, based on predicted transit depth and probability, for archival searches and for guiding ground-based follow-up observations hunting for hidden planets.

\end{abstract}


%% and \citet commands to identify citations.   The citations are
%% tied to the reference list via symbolic KEYs.  The KEY corresponds

\section{Introduction} \label{sec:intro}

Data from the \textit{Kepler Space Telescope} mission \citep{bor10} have shown that multi-planet systems are intrinsically common \citep[e.g.,][]{lis12, row14}.  However, due to the overall low detection efficiency of extrasolar planets, our understanding of the census of planets as well as the planetary architecture in any given system remain highly incomplete \citep[e.g.,][]{bat14, bur15}.  Nevertheless, methods considering selection biases due to exoplanet detection, vetting, and confirmation \citep[e.g.,][hereafter M18]{dre15, mul15, fre13, mul18} provide a robust basis to understand planetary architectures statistically. It is now becoming possible to use this population-level knowledge to predict yet-undetected planets in systems with multiple planets. In this paper we present a framework for such predictions.

Successful predictions of yet-undetected but potentially detectable planets are important for at least four reasons: First, with high-precision observations typically limited by resource availability (e.g., telescope time), prioritizing targets for follow-up observations is important for maximizing yield and efficiency \citep{qui19}.

Second, by increasing the number of known planets in multi-planet systems, observations guided by a successful prediction method can help complete our understanding of planetary architectures.  At the same time, model-driven searches for predicted planets will also directly test different possible descriptions of planetary architectures and inform our population-level understanding of orbital architectures of planetary systems \citep[][hereafter H19]{hef19}.

Third, key elements of our understanding of exoplanets come from understanding how they formed.  With the discovery that most exoplanetary systems are different from the Solar System \citep[e.g.,][]{mul19}, multiple planet formation pathways have been proposed, including different contributions from newly-recognized processes \citep[][]{mor18, ems20}.  A search for planets driven by a prediction method that is based on models of planet formation pathways will also test the planet formation models. 

Fourth, finally, it remains very difficult to detect potentially habitable extrasolar planets, and such detections typically require major, concentrated effort from larger collaborations \citep[e.g.,][]{ang16}.  A successful prediction method could play a major role in guiding the small number of intensive campaigns that are required to detect habitable exoplanets.  Also, if the prediction is based on general understanding of the formation of the planetary system \citep[e.g.,][]{wan20}, it could constrain several of the key observable (albeit not directly) parameters that are required to assess the potential habitability of the targeted planet \citep[][]{apa19a}.

It is now conceivable to develop such a successful planet prediction framework because, over the past five years, our population-level understanding of planetary architectures has improved significantly. For example, we now know that stars tend to contain at least one planet \citep{you11, fre13}.  Multi-planet systems tend to have similar orbital architectures; planets are either found with similar spacing in log period between pairs \citepalias{mul18}, or in some cases in clusters where the central period can depend on the planets' physical characteristics \citepalias{hef19}.  Planets are not always found along integer resonances, but have a peaked distribution in their pairwise planet period ratios \citep{fab14, ste15}.  In addition, these systems tend to be dynamically packed near the stability limit \citep[e.g.,][]{lis11, fan12, mal15, vol20}; dynamical instabilities also constrain planetary architectures \citep{wuz19}.  Viewing multi-planet systems through a set of more global system parameters shows most multi-planet systems are likely to be less complex than can be expected from statistical models \citep{gil20}.  Therefore, systems with higher complexity in planet spacing and gaps might have additional unseen planets.

Although our knowledge base on multi-planet systems is large and continues to grow, observational biases are a known limitation.  Population synthesis models extrapolate the current data beyond these limits to generate a representative sample of exoplanet systems.  The Exoplanet Population Observation Simulator (EPOS, \citetalias{mul18}) and the Exoplanets Systems Simulator (SysSim, \citetalias{hef19}) are examples of forward models that take the ensemble of exoplanet system statistics and simulate new exoplanet populations.  They take into account observational biases and detection efficiencies to fine-tune the parameters such that the detectable portion of the suite of simulated planets matches the observed \textit{Kepler} sample.  The differences in the base distributions provided from these simulators can be tested by conditioning on data from already known planets and making predictions on additional hidden planets.

Model predictions can be tested via follow-up observations of these previously known systems.  The \textit{Transiting Exoplanet Survey Satellite} (\textit{TESS}, \citeauthor{ric15} \citeyear{ric15}) has been operating in its nominal mission since 2018 and has found dozens of multi-planet systems.  Organized continuous observations of these systems via the Exoplanet Follow-up Observation Program for \textit{TESS}\footnote{\url{https://exofop.ipac.caltech.edu/tess}} \citep[ExoFOP-TESS,][]{qui19} confirm planet candidates and continue the search for more planets in these systems.  There is a large parameter space for each multi-planet system in which currently unknown planets could be discovered.  Predictive models can direct follow-up observation programs by determining which of these targets are more likely to contain an additional planet.

The goal of this manuscript is to present a planet prediction framework. Here, we introduce the DYNAmical Multi-planet Injection TEster (\mname{})\footnote{\url{https://github.com/JeremyDietrich/dynamite}}, a model to predict the presence of unseen planets in multi-planet systems via population statistics, as well as applications of \mname{} for the \textit{TESS} multi-planet systems.  In Section \ref{sec:methods} we lay out the general formalism for \mname{}, state our overarching assumptions, and provide the specific statistical implementation used to generate our results; the implementation is extendable as statistical knowledge improves.  We perform a sensitivity assessment to verify the robustness of \mname{} in Section \ref{sec:sav} and show the applications to \textit{TESS} multi-planet systems in Section \ref{sec:apps}.  The results from our analysis on the \textit{TESS} sample are shown in Section \ref{sec:results}, and we discuss in Section \ref{sec:discussion} the implications from these results for the \textit{TESS} systems, the different \mname{} model parameters, and the verification and confirmation of planet candidate transit signals.

\section{Methods} \label{sec:methods}

\subsection{General approach}

Our goal is to determine the likelihood of finding another currently unknown planet in a multi-planet system using planetary statistics derived from large populations.  The formalism of this likelihood is a triple integral of the probability density function over the inclination, period, and planet radius space.
\begin{equation}
    L_p = \int_0^{180} \int_{R_{min}}^{R_{max}} \int_{P_{min}}^{P_{max}} \Pr(P,R,i)\,dP\,dR\,di
\end{equation}
Each of these probability density functions describes how likely it is a planet exists with those parameters.  For example, $\Pr(P)$ would be the probability density function for finding a yet-unknown planet in a system with period $P$, etc.  We make the assumption that the probability function $\Pr(P,R,i)$ is separable in the three dimensions; that is, the probability distributions w.r.t. period, planet radius, and inclination are not dependent on the other two parameters \citepalias{mul18}.  We explore and justify this assumption in Section \ref{sec:just}.  Thus,
\begin{equation}
    \Pr(P,R,i) = \Pr(P) \Pr(R) \Pr(i),
\end{equation}
and we will now consider the three components of the likelihood function separately. The period distribution for a system with known planets is:
\begin{equation}
    \Pr(P) = \left\{
    \begin{array}{ll}
        1, & P_j = P_k \textrm{ and } j = k \\
        0, & P_j = P_k \textrm{ and } j \neq k \\
        f(P), & P_j \neq P_k \textrm{ and } P_{min} < P_j < P_{max} \\
    \end{array}
    \right.
\end{equation}
where $P_j$ is the period of planet $j$ in the system (known or hypothetical), $P_k$ is the orbital period of known planet $k$ in the system, $f(P)$ is defined by a probability density function (PDF), and $P_{min},\,P_{max}$ are the limits on the period.  The probability of finding another planet at the orbital period of a known planet is 0, since there's no stable configuration where this can occur.

The probability distribution for the planetary radius is:
\begin{equation}
    \Pr(R) = \left\{
    \begin{array}{ll}
        1, & R_j = R_k \textrm{ and } j = k\\
        f(R), & (R_j = R_k \textrm{ and } j \neq k) \textrm{ or } (R_j \neq R_k), \\
        & R_{min} < R < R_{max}\\
    \end{array}
    \right.
\end{equation}
where $R_j$ is the planetary radius of planet $j$, $R_k$ is the radius of known planet $k$, $f(R)$ is defined by a PDF of planet sizes and $R_{min},\,R_{max}$ are the planet radius limits.  The posterior probability of finding a planet with radius $R_k$ is 1, since we know at least one planet in said system has that planetary radius, but the probability of finding another planet with that radius is determined from the PDF as there are no exclusionary limits on the planet radius.

Similarly, the probability distribution function for the inclination is:
\begin{equation}
    \Pr(i) = \left\{
    \begin{array}{ll}
        1, & i_j = i_k \textrm{ and } j = k\\
        f(i), & (i_j = i_k \textrm{ and } j \neq k) \textrm{ or } (i_j \neq i_k), \\
        & 0 < i < 180\\
    \end{array}
    \right.
\end{equation}
where $i_j$ is the inclination of planet $j$, $i_k$ is the inclination of known planet $k$, and $f(i)$ is defined by a PDF of planet inclinations.  The inclinations are drawn from 0$^\circ$ (face-on prograde) to 180$^\circ$ (face-on retrograde), with 90$^\circ$ being edge-on.  The system inclination, which is the plane on which the orbits are centered, is close to 90$^\circ$ for systems with transiting planets.  The mutual inclinations between planets in the system are drawn from a Rayleigh distribution with a relatively small scatter centered around the system inclination, which is a parameter to the function that can be inferred from fitting the known planet inclination distribution \citep{fan12, fab14}.

Once the probability function is determined across all three dimensions, we sample it via the Monte Carlo method. We take $N = 10,000$ iterations of each system and inject planets with periods, radii, and inclinations determined by sampling the probability function.  The systems with the highest relative likelihood in the shortest periods can then be prioritized for ground-based follow-up observations.

\subsection{Correlations in Model Parameters}\label{sec:just}

An approximation made in the current version of \mname{} is that the period, planet radius, and orbital inclination of the planet are independent of each other.  In the following we will explore the justification for this approximation.

Multiple works \citep[e.g.,][]{hel16, car18} have shown that short-period planets tend to be smaller than planets further out, as the host star evaporates any light gas envelope.  This trend is most significant for large planets ($R > 4-5 R_\oplus$; \citeauthor{cia13} \citeyear{cia13}, \citeauthor{hua16} \citeyear{hua16}), which we exclude from our sample. \citet{wei18} and \citetalias{mul18} show that for Kepler targets, planets within the same system tend to be the same size compared to random draws. \citet{wei18} found a correlation between pairwise inner and outer planet radii, but \citet{gil20} showed this correlation is unclear and may be due to scatter within low-monotonicity systems.  \citet{daw16} showed that flatter systems with smaller mutual inclinations tend to have smaller planets with closer orbital spacing, whereas systems with larger mutual inclinations tend to include gas giants with larger orbital spacings.  \citet{gil20} found that the mass scale (and thereby the radius scale) of a system does not correlate to the flatness, but does reproduce the slightly weak but significant correlation of tighter spacing in flatter systems.

Although evidence exists for correlations between fundamental planet parameters, given current data these correlations are relatively weak, and the mechanisms behind these correlations are not yet fully understood.  Given that planet properties within a system are only weakly correlated and that there is no well-established, quantified correlation function we could implement, in our study we take the assumption that the parameters are independent. This, we stress, is an approximation, one which can be tested by the predictions given out by \mname{} and refined when a more accurate population model can quantify these correlations across all the variables.  \mname{} would then be updated to include these correlations in the formalism.

\subsection{Assumptions}

We focus our study on the orbital period range from 0.5 to 730 days, for which exoplanet population data is available (see, e.g., \citetalias{mul18}).  The occurrence rates of planets are not well defined interior to 0.5 days and beyond 2 years.  It is extremely difficult to find long-period transiting planets due to the amount of time needed as well as the low probability of having the correct inclination to transit; as the simple geometric probability of transiting goes as $R_*/a$, the inclination limits for a transiting planet are closer to 90$^\circ$ for a planet at a larger semi-major axis $a$.

In this study we assume planet radius limits from 0.5 to 5 $R_\oplus$, as a large majority of the \textit{TESS} multi-planet systems have planet radii that fall between these boundaries.  It has been noted that planets with radii above 3-4 $R_\oplus$ are below an observed ``break" in the planet radii distribution across all systems, and thus may have their own statistical distribution (\citetalias{hef19} and references therein).  The \textit{Kepler} sample does not constrain populations below 0.5 $R_\oplus$, as they are often too difficult to detect en masse.  Smaller, rocky (super)Earth-like planets are more common in the galaxy than Jupiter-like gas giant planets within 1 AU \citep[e.g.,][]{how12, mul15}, and finding them were direct mission goals for \textit{TESS} and \textit{Kepler}, as well as ground-based follow-up observations that are still ongoing and the main reason we prioritize systems likely to contain these planets in our statistical analysis.

We assume the inclination of a planet can be any value between 0$^\circ$ and 180$^\circ$ but is constrained by the PDF of our distribution.  However, since the measurement of the inclination is degenerate (i.e., we receive the same signal for planets with inclination $i$ and $180 - i$), we assume the inclination of the first planet is $<90$ degrees.  Also, for a high number of known transiting planets ($N \geq 4$), we assume the planet inclinations follow the Rayleigh distribution only, whereas for $N = 2, 3$ we assume an intrinsic scatter where some fraction of systems have inclinations distributed isotropically as per the ``\textit{Kepler} dichotomy" (\citeauthor{joh12} \citeyear{joh12}, \citetalias{mul18}) but where the fraction is scaled downwards from the isotropic fraction of singly-transiting planets.  For example, if the isotropic fraction of systems with one known planet is 30\%, then the isotropic fraction of systems with two planets is 20\% and three planets 10\%.  This is done so as to not have a large break in the inclination distribution between singly-transiting systems and doubly-transiting systems if the isotropic fraction is high; the odds of a misaligned planet in a doubly-transiting system should be higher than a misaligned planet in a 4-transiting-planet system \citep{bal16}.

To ensure that the new systems created after injections are dynamically stable on long timescales, we use the dynamical stability parameter $\Delta$, which is defined as
\begin{equation}
\Delta = \frac{2(a_2 - a_1)}{a_2 + a_1}\left(\frac{3M_*}{m_2 + m_1}\right)^{1/3},
\end{equation}
where $a_1, a_2$ are the semi-major axes for the orbits of the inner and outer planet respectively, $m_1, m_2$ are the inner and outer planet masses, and $M_*$ is the stellar mass.  The innermost theoretical limit for $\Delta$ is $\Delta_c = 2\sqrt{3} \approx 3.46$ \citep{gla93}, but many studies have shown that long-term stability requires $\Delta_c \gtrsim 10$ for systems of three or more planets with similar masses/sizes, like those seen in the \textit{Kepler} dataset \citep{cha96, fan12, puw15}.

For our minimum threshold for planet dynamical stability, we take the more inclusive value $\Delta_c = 8$ from \citetalias{hef19} in order to allow for a larger area of period space to be tested, as some known systems have smaller $\Delta$ values (e.g., the two planets of Kepler-36; \citeauthor{car12} \citeyear{car12}).  In most cases for planets discovered via the transit method, the planet masses are unknown, so we use the non-parametric mass-radius relation from \citet{nin18} to formulate a range of masses corresponding to the range of planet radii for the injected planets.

\subsection{Implementation} \label{subsec:implement}

We defined two separate period PDFs to study and understand the impact of the assumption on the results of fitting current systems to different model distributions and testing the resulting likelihoods.  We first used the period ratio model from \citetalias{mul18}.  The period ratio model assumes a broken power-law for the location in period space of the first planet in the system with the break at 12 days, where the peak is in the first-planet statistics from \textit{Kepler} \citepalias{mul18}.  Outwards from the first planet, it calculates a dimensionless spacing between consecutive planets $D_k$, defined as 
\begin{equation}
    D_k = 2\frac{\mathcal{P}_k^{2/3} - 1}{\mathcal{P}_k^{2/3} + 1},
\end{equation}
where $\mathcal{P}_k = P_{i+1}/P_k$.  The period ratio model assumes $D_k$ follows a lognormal distribution with parameters $\mu_D$ and $\sigma_D$.  Thus, the full period distribution is
\begin{equation}
    f(P) = \left\{
    \begin{array}{ll}
        1, & P = P_j\\
        (P/P_{break})^{a_P}, & p_k = 1 \textrm{, } P < P_{break},\\ & \textrm{and } P < P_1\\
        (P/P_{break})^{b_P}, & p_k = 1 \textrm{, } P > P_{break},\\ & \textrm{and } P < P_1\\
        \frac{1}{\sqrt{2\pi\sigma_D^2}}e^{-\frac{(\log D_k - \mu_D)^2}{2\sigma_D^2}}, & p_k > 1\\
        0, & \Delta_{jk} < \Delta_c\\
    \end{array}
    \right.
    ,
\end{equation}
where $P_j$ is the period of the $j$th known planet in the system (so $P_1$ is the period of the innermost known planet), $p_k$ is the position of a new planet counting outwards from the star, $a_P$ and $b_P$ are the power-law parameters, and $\Delta_{jk}$ is the stability criterion measured between known planet $j$ and inserted planet $k$.  Values for $\mu_D$, $\sigma_D$, $a_P$, and $b_P$ are found in Table \ref{tab:params}.  In particular, this model tends to insert planets roughly symmetrically in gaps, as it is not sensitive to the planet masses. However, the stability criterion relies on the masses and cuts off the period probability distribution at a further distance from a more massive planet.

Secondly, we used the clustered periods model from \citetalias{hef19}.  To determine the cluster periods (the locations in period space where known planets in a system tend to be found), the clustered period model iterates through a set of test periods and determines the fit of the known planetary periods in the system (scaled by the current test cluster period), to a lognormal distribution,
\begin{equation}
    f(P_i) \propto P_c\times\textrm{Lognormal}(0, N_P\,\sigma_P),
\end{equation}
with $N_p$ being the number of planets in the cluster and $\sigma_P = 0.2$ \citepalias{hef19}.  If planets in the system have scaled periods beyond the 97th percentile of the lognormal distribution that fits the remaining planets best (a variation of $\sim 2.5\sigma$), we assume a two-cluster model where the probability of finding a planet at a certain period can be in one of two clusters.  This can be extended out beyond two clusters, but for almost all systems in the \textit{TESS} multi-planet catalog the two-cluster model is sufficiently complex to describe the orbital hierarchies, as most of the systems only contain two planets.  Thus, our full period distribution for the clustered periods model is
\begin{equation}
    f(P) = \left\{
        \begin{array}{ll}
        1, & P = P_j\\
        P_{c,i}\frac{1}{\sqrt{2\pi\sigma^2}}e^{-\frac{(\log P)^2}{2\sigma^2}}, & P/P_{c,i} < 2.5\sigma\\
         & \textrm{ and }\sigma = (N_P)_i\sigma_P\\
        0, & \Delta_{jk} < \Delta_c\\
    \end{array}
    \right.
    ,
\end{equation}
where $P_{c,i}$ is the cluster period for the $i$th cluster (clusters separated by 2.5$\sigma$), $(N_P)_i$ is the number of planets in that cluster, and $\Delta_{jk}$ is the stability criterion measured as before.  This model tends to insert planets close to the stability criterion and is more sensitive to the masses of the known and injected planets, as the injected planets are closer in period-space to the lower-mass planets.  \mname{} is most effective at calculating the likelihood of planets interior to the outermost known planet in a system, as actual planet occurrence is unknown at long periods. \mname{} prioritizes systems with gaps and large likelihoods for planets in those gaps.

For our planet radius distribution, we use a variation on the clustered periods and sizes model from \citetalias{hef19}.  Here, the clusters are locations in planet radius space where known planets in a system have similar values.  We fit all the known planetary radii in the system to lognormal distributions for a range of cluster radii, 
\begin{equation}
    f(R_i) \propto \textrm{Lognormal}(R_c, \sigma_R),
\end{equation}
with $\sigma_R = 0.3$ \citepalias{hef19} and finding the cluster radius or radii where the sets of known planets fit best.  The number of clusters is determined by the 97th percentile of the lognormal distribution of the first best-fit cluster; if the planetary radius of a known planet in the system is beyond this value, we conclude that the planet is in a separate cluster and adjust our PDF accordingly.  We do not assume that periods and planet radii are clustered together, as that would invalidate our assumption that all three parameters are independent.

Inclinations are taken from a broken distribution similar to \citetalias{mul18}.  In this distribution, 
\begin{equation}
    f_{iso} = 0.38 \times \frac{4 - x}{3},
\end{equation}
where $x = \textrm{min}(4, N)$ and $N$ is the number of planets in the system.  A fraction $f_{iso}$ of systems have inclinations distributed isotropically (and when $x = 1$, $f_{iso} = 38\%$; \citetalias{mul18}), otherwise they are chosen from a Rayleigh distribution centered on the system inclination $i_s$.
\begin{equation}
    f(i_i) \propto \left\{
    \begin{array}{ll}
        i_s + \textrm{Rayleigh}(\sigma_i), & i_{iso} > f_{iso} \\
        \sin(i_i), & i_{iso} \leq f_{iso} \\
    \end{array}
    \right.
\end{equation}
with scale factor $\sigma_i = 2$ \citep{fab14}.

The system inclination $i_s$ is determined a similar way to the cluster periods and radii, by fitting a Rayleigh distribution with scale factor $\sigma_i$ to the known planet inclinations.  After constraining the first planet inclination to $<90$ degrees, we permute all the remaining planets across the degeneracy.  Whichever permutation of planet inclinations that fit the Rayleigh distribution best for a given $i_s$ defines that as the system inclination.

We determine the dynamical stability threshold $\Delta_c$ for the known planets in the system compared to an average mass planet for that system (either calculated from the planetary mass values or taken from the average planet radius and the mass-radius relation), and we set the probability of finding a planet at the corresponding periods where $\Delta_c < 8$ to 0. A full list of parameters in \mname{} and the values assigned to them can be found in Table \ref{tab:params}.

\begin{table*}[ht]
    {\centering
    \caption{\mname{} Parameters}
    \label{tab:params}
    \begin{tabular}{|l|c|c|r|}
        \hline
        Parameter Description & Symbol & Value & Reference \\
        \hline
        Dimensionless spacing parameter mean & $\mu$ & -0.9 & \citetalias{mul18} \\
        Dimensionless spacing parameter scatter & $\sigma_D$ & 0.4 & \citetalias{mul18} \\
        Planet radius cluster scatter & $\sigma_R$ & 0.3 & \citetalias{hef19} \\
        Fraction of systems with isotropic inclinations & $f_{iso}$ & 0.38 & \citetalias{mul18} \\
        Inclination Rayleigh distribution scatter & $\sigma_i$ & 2 & \citet{fab14} \\
        Dynamical stability threshold & $\Delta_c$ & 8 & \citetalias{hef19} \\
        \hline
    \end{tabular}
    }
    \\[10pt]
    \textbf{Notes}: The set of parameters used to define our implementation of the \mname{} formalism to constrain the likelihood of finding additional planets in known multi-planet systems, with their description, symbol, value, and reference.
    \\[10pt]
\end{table*}

\section{Sensitivity Assessment and Verification} \label{sec:sav}

We tested the sensitivity and robustness of \mname{} to certain alterations in multi-planet system architectures in order to ensure its accuracy and predictive power.  In each of these tests, we changed the structure of the multi-planet system used in the test by removing one or more planets or planet candidates from the system entirely, treating the system as if that planet or candidate was not known to exist.  We then ran \mname{} to see what it would predict at the period(s) where the removed planet or planets had been located, to test that its predictions are sensitive and robust.  In one case we also tested perturbing the orbital periods of the remaining planets in the system, to see if \mname{} was robust to small uncertainties.  We expected to see a high relative likelihood of a planet existing at or near the periods where we removed planets, and the results for the most part agreed.  In all cases, we also looked for any difference in planetary radius and orbital inclination caused by the system alteration, and the distributions skewed towards the new means as expected.

\subsection{Removing a known transiting planet/reproducing a known non-transiting planet}

\begin{figure}[t]
    \centering
    \includegraphics[width=\columnwidth]{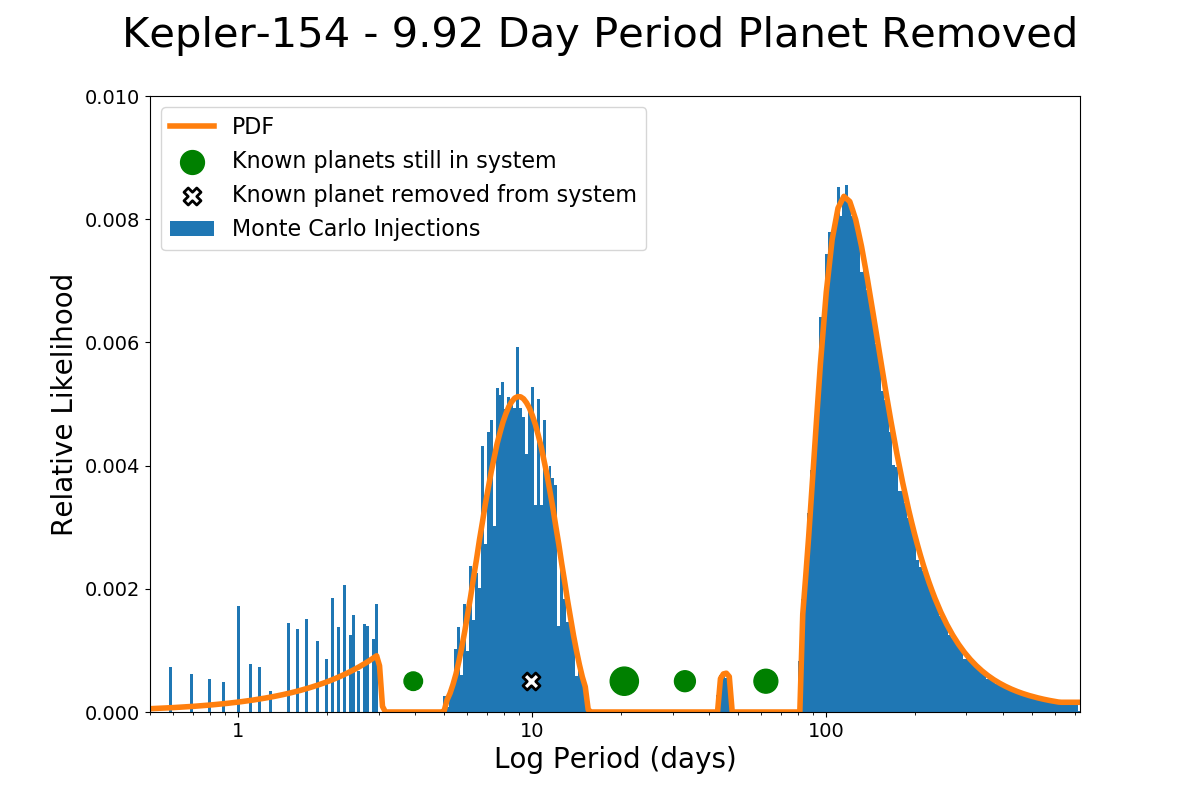}
    \caption{The Kepler-154 system without the 3.4-year period gas giant planet and with the 9.92-day period planet Kepler-154 f marked as removed.  Marker sizes correspond to the relative sizes of the planets \citep{row14, mor16}.  Almost 97\% of the Monte Carlo iterations where a planet was injected interior to the outermost planet are in the gap where Kepler-154 f was removed, whereas the other areas integrate to $\sim$3\%.}
    \label{fig:kep154}
\end{figure}

First, we tested removing a known transiting planet from a system, choosing systems with high multiplicity as they provide the most detailed prior information for our predictions.  \object{Kepler-154} is a 6-planet system with 5 smaller planets in a short-period orbital chain (periods from 4 to 62 days) and a gas giant much further out \citep[$P \sim$ 3.4 years;][]{row14, mor16, kaw19}. We run the \mname{} on the inner transiting planets of Kepler-154 (i.e., assuming the presence of 3.4-year-period gas giant does not affect the interior planets and removing it from the system).  Also, we exclude the last verified candidate KOI-435.06 = Kepler-154 f with a period of 9.92 days, as it was the transiting candidate with the lowest confidence/highest false alarm probability \citep[see Figure \ref{fig:kep154}]{mor16}. 
When running \mname{} on the four remaining planets, we find that a large percentage (96.7\%) of injected planets interior to the outermost planet are found between the first and second planets (Kepler-154 e at 3.93 days and Kepler-154 d at 20.55 days; uncertainties on the periods measured from \textit{Kepler} are $< 10^{-4}$ days).  90.1\% of interior injections occur within 5 days of 9.92 days, with a mode at 9.9 days that is consistent with the planet period within the fineness of our model.  The radius of Kepler-154 f is measured as $1.50^{+0.34}_{-0.21}R_\oplus$, and 67\% of \mname{}'s predicted planets have a physical radius within 3$\sigma$ of the known radius.

\begin{figure}
    \centering
    \includegraphics[width=\columnwidth]{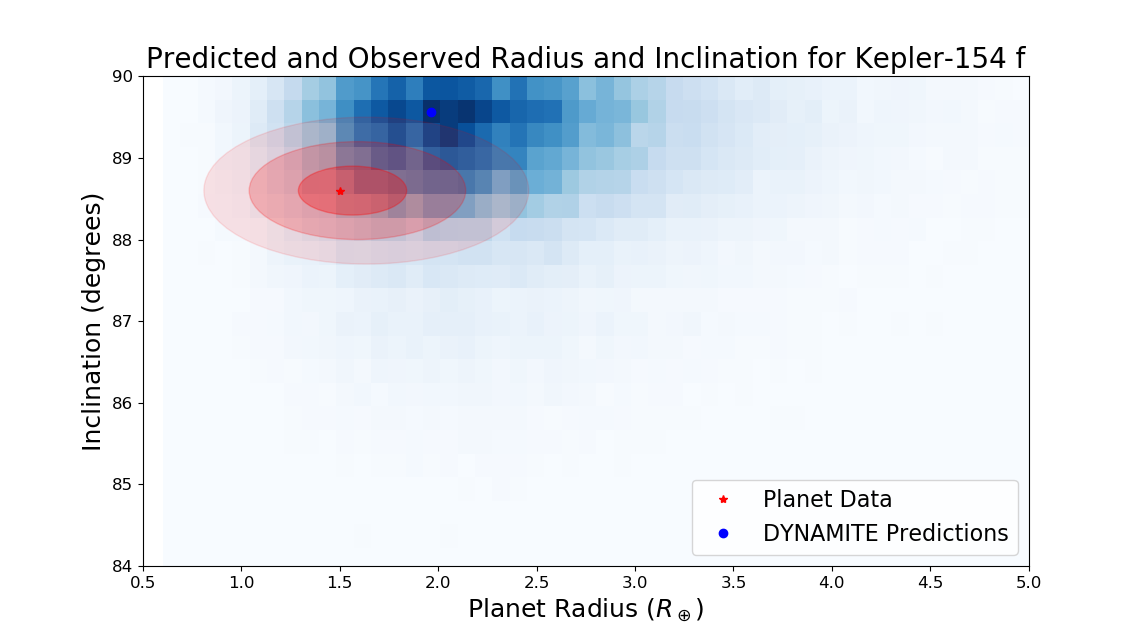}
    \caption{A comparison of the known data for Kepler-154 f and the corresponding predictions from \mname{}.  $1\sigma$, $2\sigma$, and $3\sigma$ uncertainties for the data are shown as ellipses.  The predictions show good agreement with the known data, to the level at which it can inform future observations of the target.}
    \label{fig:kep154_RI}
\end{figure}

The orbital inclination of Kepler-154 f is not fully constrained, but using the average impact parameter at its period of 9.92 days places it at an inclination of $88.6 \pm 0.2^\circ$, which is near the calculated plane of the system (as the remaining transiting rocky planets have inclinations of $85.0 \pm 0.2^\circ$, $88.5 \pm 0.3^\circ$, $89.5 \pm 0.5^\circ$, and $89.4 \pm 0.4^\circ$, as calculated from their impact parameters in \citet{mor16}).  \mname{} predicts $88.9 \pm 0.8 ^\circ$ as the orbital inclination of the injected planet, matching the predicted inclination well but with a larger spread, as 43\% of the iterations from the model are within 3$\sigma$ of the estimated inclination for Kepler-154 f.  As the mean injection value from \mname{} matches the estimated value of the orbital inclination very well, we find this agreement to be very strong.  Good predictions need only match the inclination within a couple of degrees, to identify whether transits are likely and to constrain the degeneracy on $M \sin i$ for RV planets.  A comparison of the known data with the predictions from \mname{} is shown in Figure \ref{fig:kep154_RI}.

\begin{figure}[t]
    \centering
    \includegraphics[width=\columnwidth]{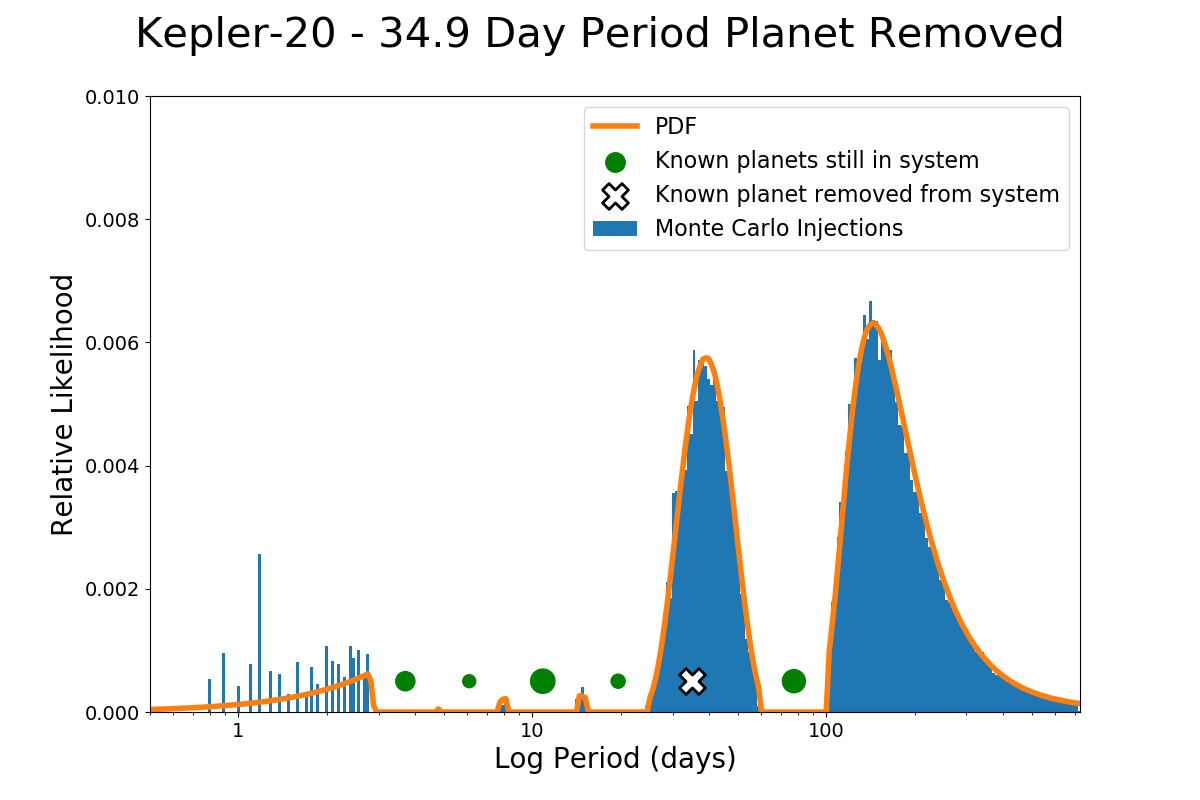}
    \caption{The Kepler-20 system with the 34.94-day period planet Kepler-20 g marked as removed.  Marker sizes correspond to the relative sizes of the planets \citep{fre12}.  The size estimate for the non-transiting planet Kepler-20 g was calculated via its mass measurement from \citet{buc16} and the mass-radius relationship from \citet{nin18}.  \mname{} predicts that if a planet should exist interior to Kepler-20 d (the outermost planet), 99.7\% of the time it would exist in the gap between Kepler-20 f and Kepler-20 d, with the other areas integrating to only 0.3\%.  The position of the planet is predicted further out for the period ratio models from \citetalias{mul18}, as depicted here.}
    \label{fig:kep20}
\end{figure}

\object{Kepler-20} is a 6-planet system with all 6 planets in an orbital chain with periods from 3.6 days to 77.6 days \citep{fre12}.  The last discovered planet was Kepler-20 g, which has a period of $34.94 \pm 0.04$ days and was found not to transit \citep{buc16}.  By removing this planet, we can test the system as it was seen only through transit light curves, or as if there were no radial velocity (RV) data confirming the existence of this planet.  We ran \mname{} on only the transiting planets of Kepler-20 (see Figure \ref{fig:kep20}).

We find that there should be a very high probability (99.7\% of total possible interior injections) of a planet in between the farthest planet (Kepler-20 d at 77.61 days) and second-farthest planet (Kepler-20 f at 19.58 days), the gap in period space where Kepler-20 g was found.  Narrowing the injections down to a 10-day window centered on Kepler-20 g's known period, we still find that 42.4\% of interior injections occur in this area, with a mode at 39.0 days.  In addition, injected planets located in this gap with the estimated planet radius for Kepler-20 g transit only 22.6\% of the time, showing that it is not an outlier because it does not transit.  Due to its non-transiting status, a precise planetary radius for Kepler-20 g has not been measured, but we infer a planet radius of $3.46^{+0.30}_{-0.25} R_\oplus$ from the non-parametric mass-radius relationship. \mname{} determined that the planet radii of this system likely follow a 2-cluster model, and predicts the planet radius of an injected planet in that cluster to be within 3$\sigma$ of the inferred planet radius 55\% of the time.

In both tests of this scenario, we find that \mname{} predicts the presence of the known transiting or non-transiting planet that was removed, and is more accurate with the period ratio model than the clustered periods model.  For Kepler-154, \mname{} predicts the location of the planet in period space almost exactly, whereas for Kepler-20, \mname{} predicts a longer period than the expected period.  However, the presence of Kepler-20 g could still be inferred, and follow-up observations (both transit light curves and precision RV) would prioritize that area in period space for its discovery.  The primary function of \mname{} is to provide prioritization for follow-up targets and not to predict the parameters exactly.  Our tests show that \mname{} predicts not only the periods but also the radii and inclinations of planets to levels that are well-suited for guiding observations.  In the future a single, combined metric based on these multiple dimensions could be created to allow for, for example, optimizing the assumptions that underpin \mname{}.

\subsection{Removing multiple planets}

In this scenario, we take a 4-planet system and remove 2 planets in various configurations.  TOI 174, also known as \object{HD\,23472}, has been confirmed to have 2 sub-Neptune-sized planets ($\sim 2 R_{\oplus}$) in $17.667^{+0.142}_{-0.095}$  day and $29.625^{+0.224}_{-0.175}$ day orbits \citep{tri19}, which are called HD 23472 b and c. \textit{TESS} has also found evidence for two possibly rocky planets ($\sim 1-1.3\textrm{ } R_{\oplus}$) interior to the known planets at 4.0 and 12.2 days, called TOI 174.04 and TOI 174.03, respectively.  Removing both TOI 174.03 and TOI 174.04 from the system, we find that all interior injections occur at orbits shorter than the period of HD 23472 b, and the mode of planet injections for the entire system (interior and exterior) occurs at the \textit{Kepler} statistical first-planet mode of 12 days \citep{mul18}, showing a very high likelihood of at least one planet interior to HD 23472 b at a 17.7 day period.

\begin{figure}[t]
    \centering
    \includegraphics[width=\columnwidth]{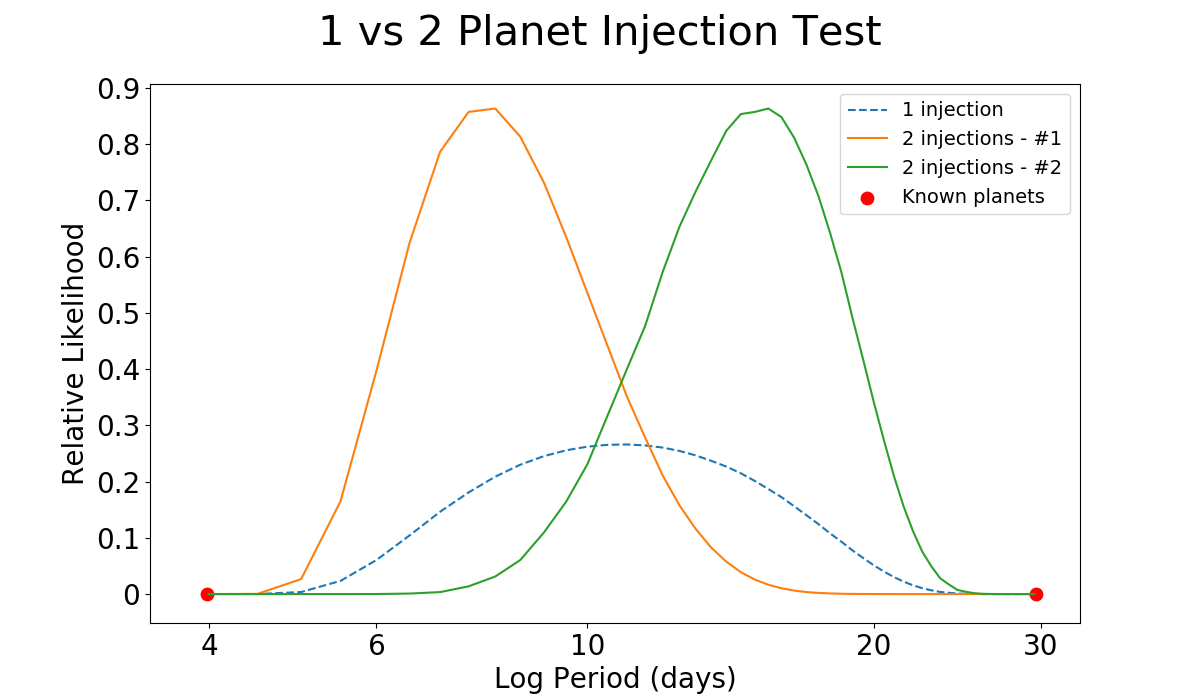}
    \caption{Testing 1 vs 2 planet injections between TOI 174.04 and HD 23472 c at periods $\sim$ 4 and $\sim$ 30 days (red dots).  The relative likelihood of two planets in that gap of period ratio $\sim$ 7.5 is much higher than the relative likelihood of one planet in that gap.}
    \label{fig:2inject}
\end{figure}

When removing TOI 174.04 and HD 23472 b, we find an almost exactly even distribution in the location of interior injections; 50.2\% are injected at orbits shorter than TOI 174.04 at 12.2 days, and 49.8\% are injected at orbits between TOI 174.04 and the remaining known planet HD 23472 c at 29.6 days, with a mode at 18.5 days. When removing TOI 174.03 and HD 23472 b, we instead find a wide distribution of planet injections with a mode of 12 days.  The total probability of injecting one planet interior to HD 23472 c is only slightly more than half of the previous case, however, suggesting that injecting multiple planets here would fit the data better (see Figure \ref{fig:2inject}).

When we remove TOI 174.03 and HD 23472 c, there is a high probability of finding two planets again, but not necessarily in the same positions as currently known.  The mode of the injections between TOI 174.04 and HD 23472 b is around 8.5 days, as two period ratios near 2 are more likely than period ratios of $\sim$3 and $\sim$1.5, respectively.  The mode of the injections exterior to HD 23472 b, following the \textit{Kepler} statistics, is $\sim$32 days, but there is still a 46\% probability of finding the outermost planet within 5 days of its current period.

\begin{figure*}[ht]
    \centering
    \includegraphics[width=0.475\textwidth]{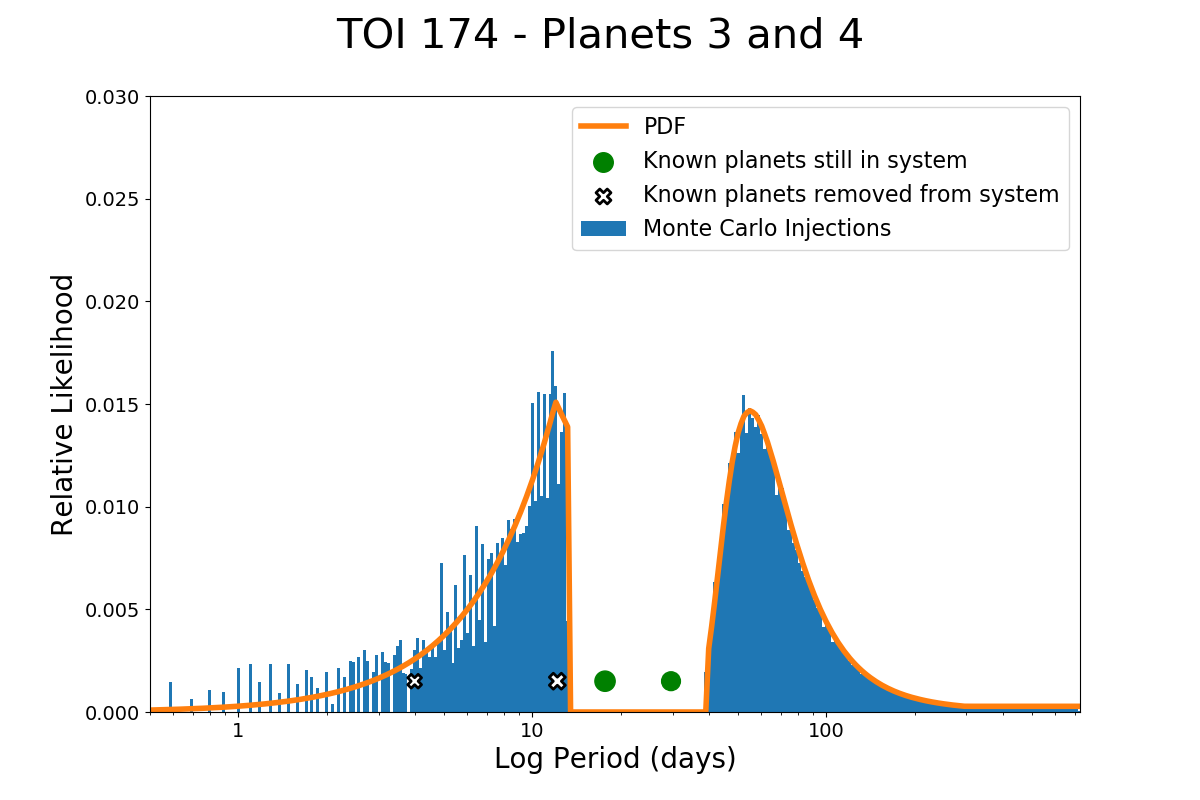}
    \includegraphics[width=0.475\textwidth]{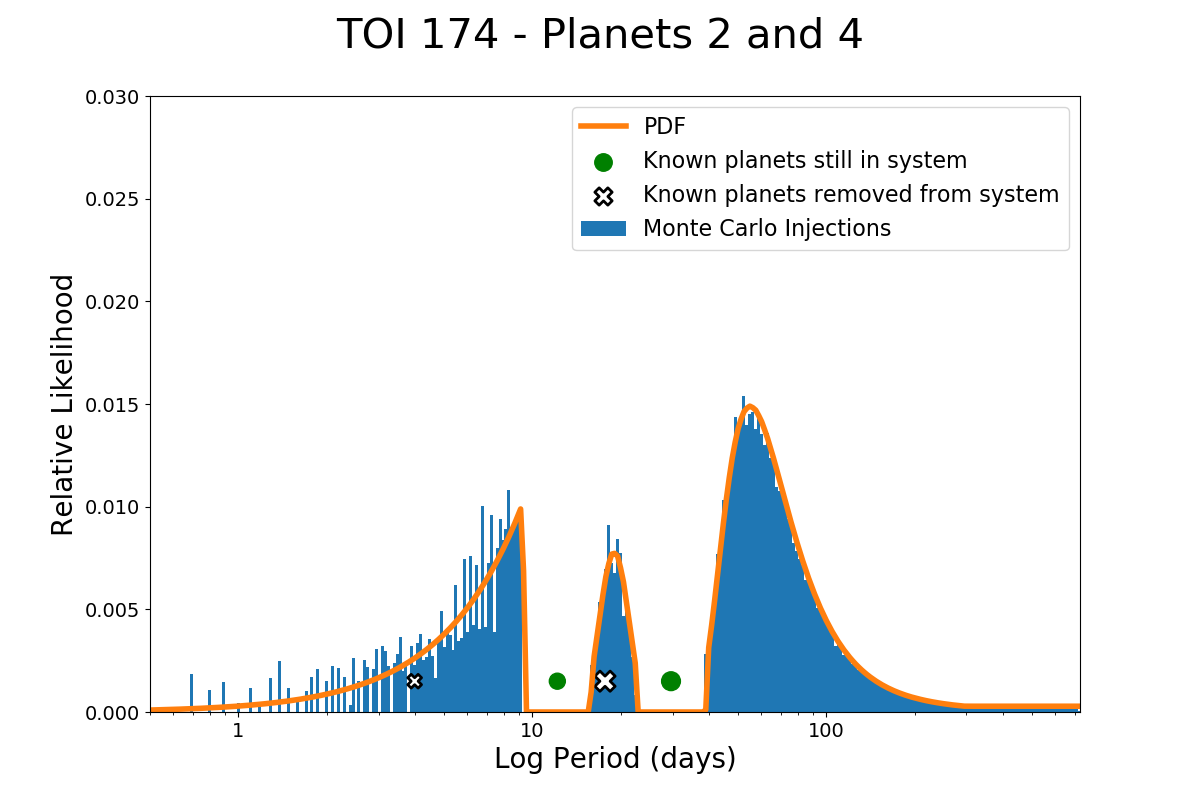}
    \includegraphics[width=0.475\textwidth]{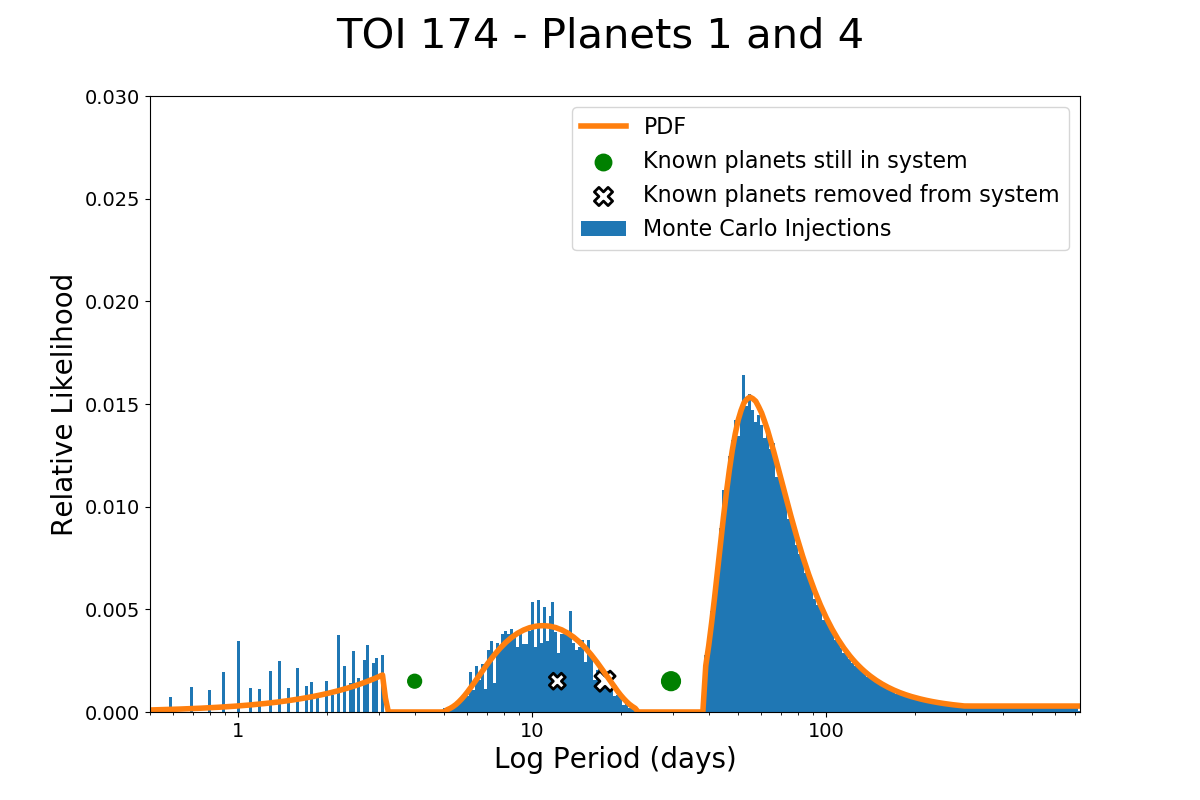}
    \includegraphics[width=0.475\textwidth]{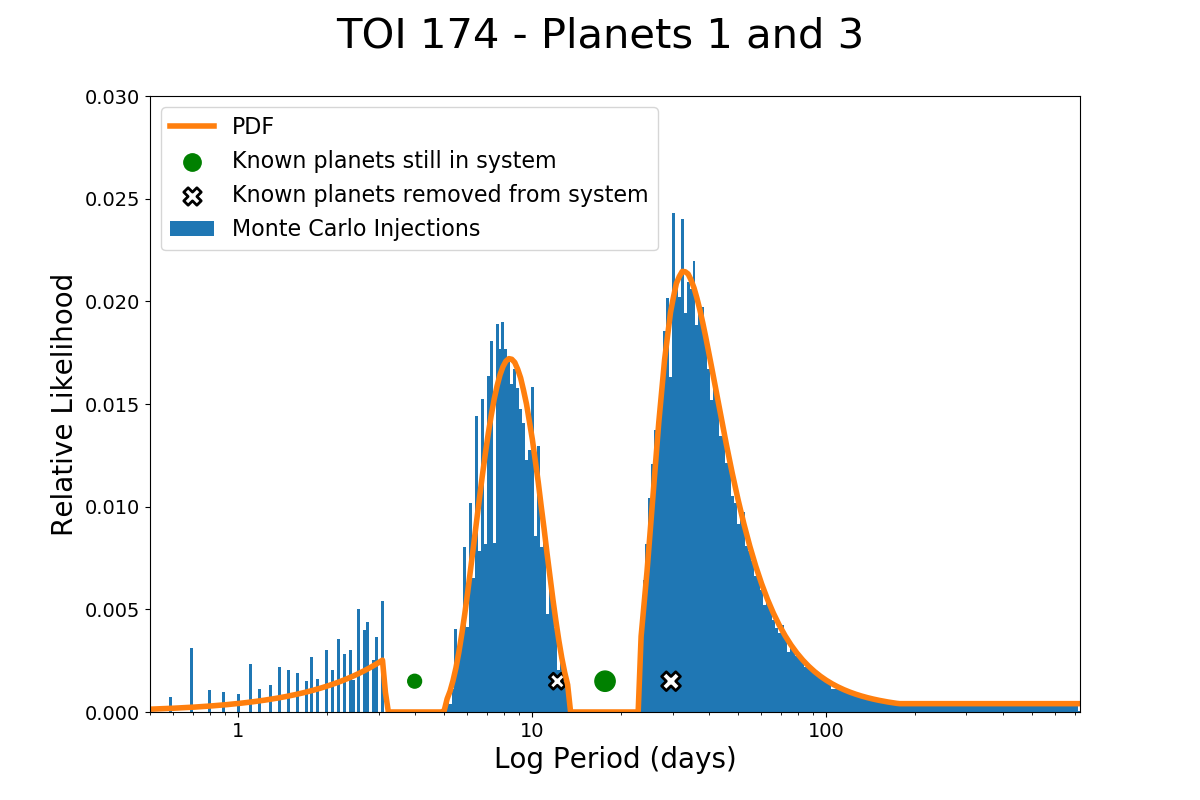}
    \caption{The TOI 174 system in various configurations. Top Left: Inner two planet candidates TOI 174.03 and TOI 174.04 were removed; \mname{} shows a high likelihood of at least one planet interior to the 17.7-day period planet HD 23472 b.  Top Right: TOI 174.04 and HD 23472 b were removed; \mname{} shows a relatively high likelihood of a planet interior to TOI 174.03 and one planet between TOI 174.03 and HD 23472 c.  Bottom Left: Planet candidate TOI 174.03 and HD 23472 b were removed; \mname{} shows a lower, more spread out likelihood of finding one planet between TOI 174.04 and HD 23472 c, indicative of two planets missing.  Bottom Right: TOI 174.03 and HD 23472 c were removed; \mname{} shows a very high likelihood of finding a planet where TOI 174.03 would be, although shifted towards lower periods due to the period ratio statistics.}
    \label{fig:toi174}
\end{figure*}

We show the different system structures after removal with relative likelihoods in log period space in Figure \ref{fig:toi174}.  In all tests of removing two out of four planets in this system, the two-cluster planet radius model failed and the two remaining objects fell into one cluster.  When removing one of the \textit{TESS} planet candidates along with one of the known planets, the cluster centers in between the two, whereas when both TOI 174.03 and TOI 174.04 are removed the cluster shifts towards the planet radii of HD 23472 b and c, and vice versa.  This would affect the prioritization to a degree, depending on what types of planets the follow-up observations are looking for.  The orbital inclinations of the planets all fell within 1$^\circ$ of each other, so removing any pair of them did not affect the plane of the system significantly.

\subsection{Removing a planet and perturbing the system}

We also tested removing one planet from the TOI 174 system and perturbing it to see if \mname{} is rigorous to changes in the system.  We altered the period of each of the remaining planets by a random amount within 3$\sigma$, as well as 5\%.  Neither perturbation altered the predictions of the mode of the injected planets, within the fineness of the model at 0.5 days; the histograms did shift slightly but insignificantly when perturbations between remaining planets occurred in the same radial direction.  The planet radius and inclination predicted values also shifted similarly.

\subsection{Analysis of planet candidates not yet verified}\label{subsec:savanv}

\begin{figure*}[ht]
    \centering
    \includegraphics[width=0.475\textwidth]{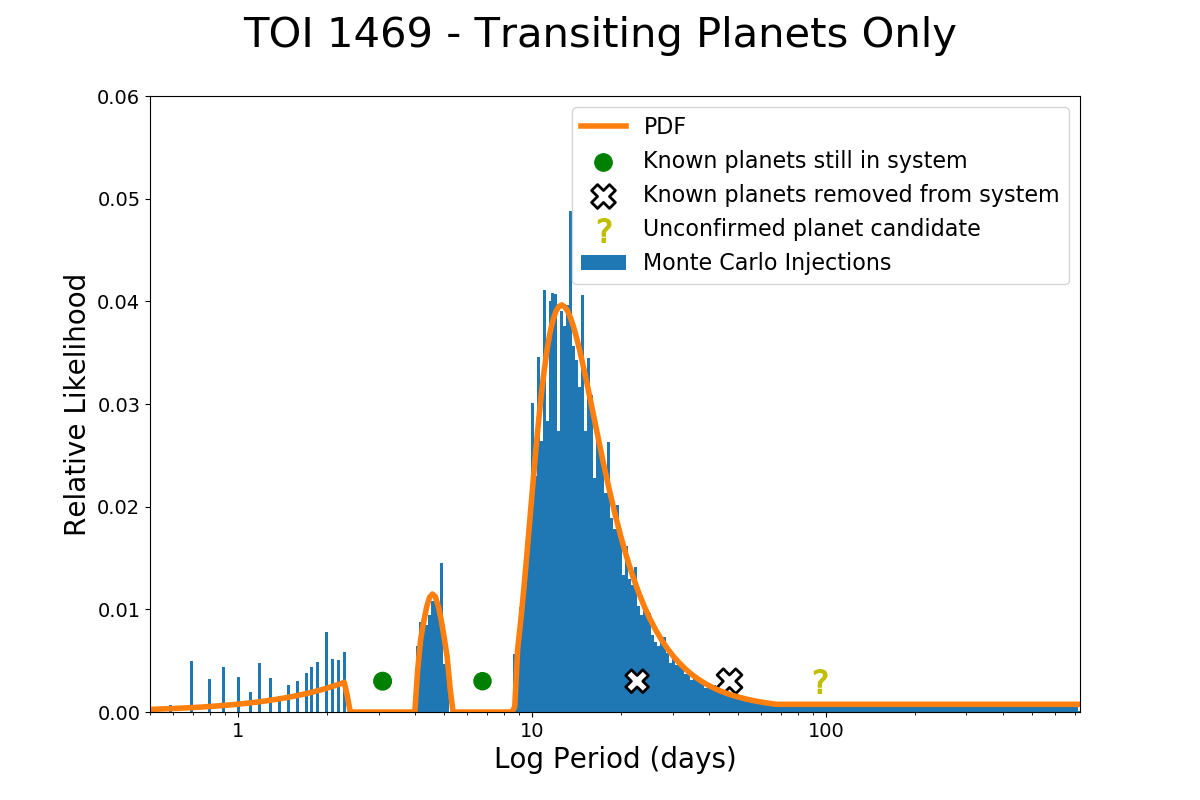}
    \includegraphics[width=0.475\textwidth]{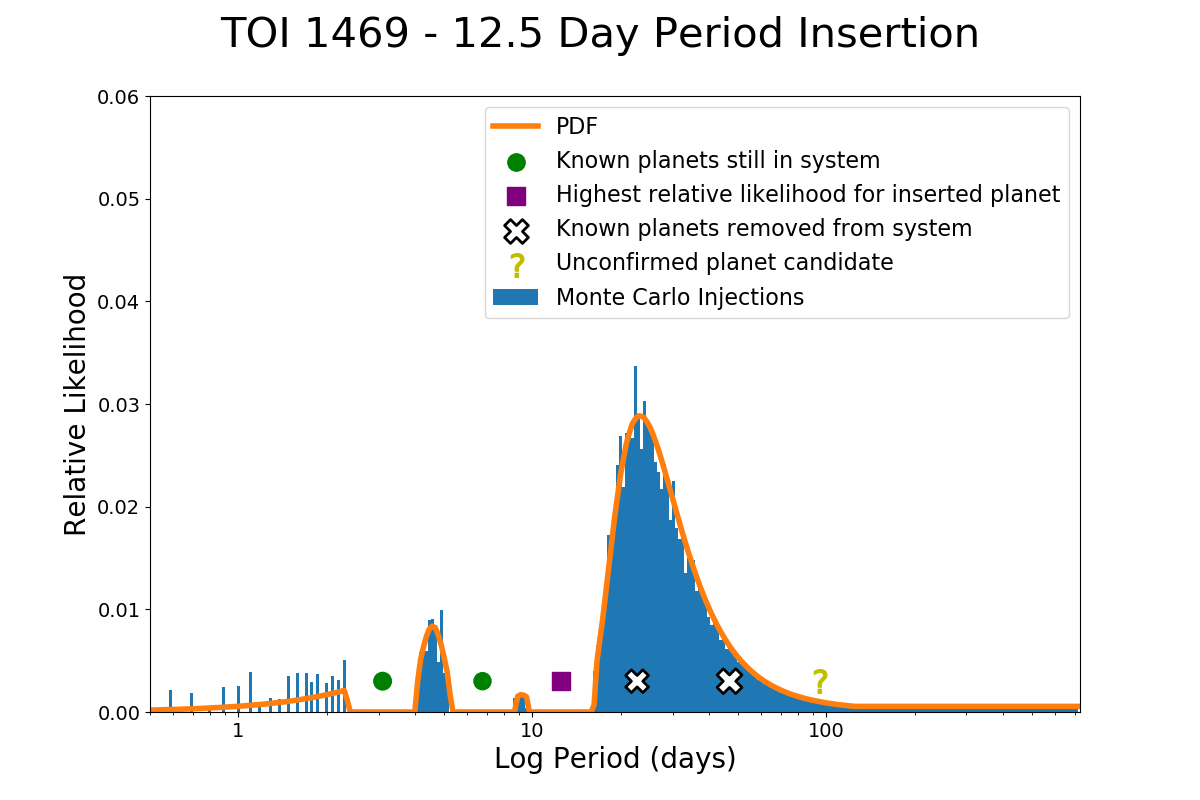}
    \includegraphics[width=0.475\textwidth]{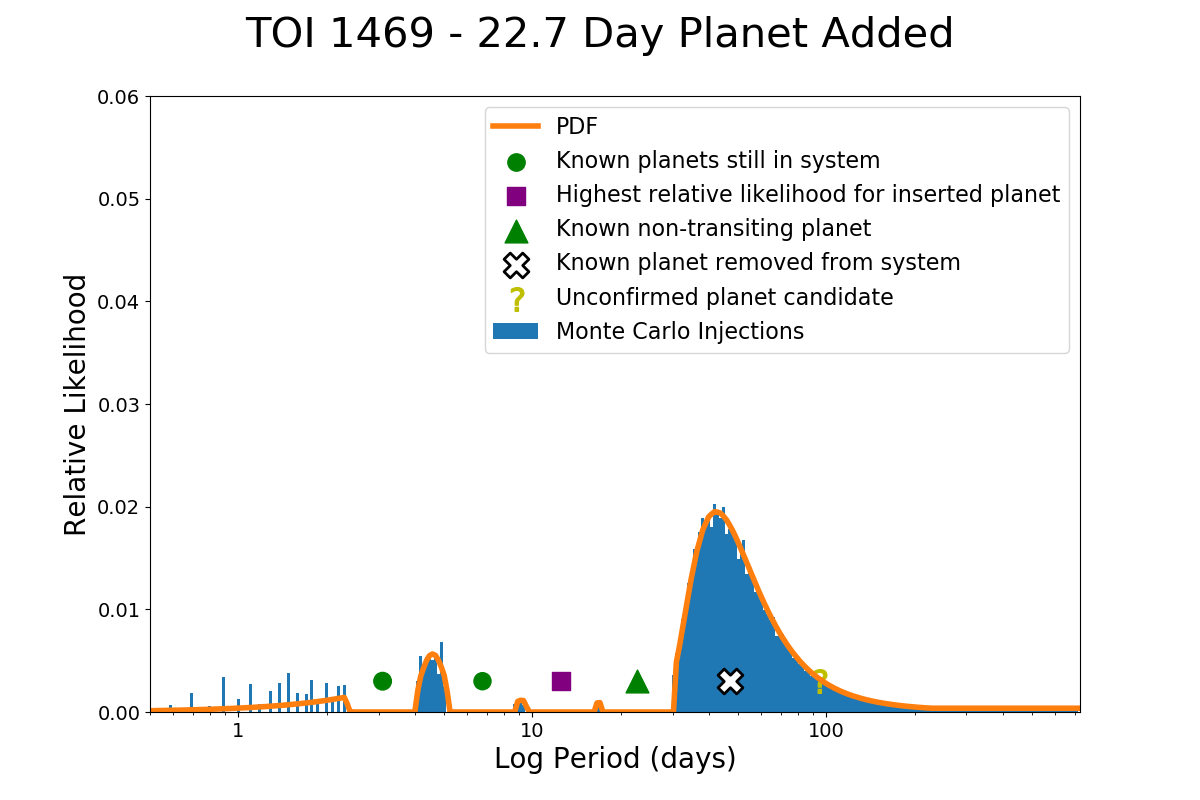}
    \includegraphics[width=0.475\textwidth]{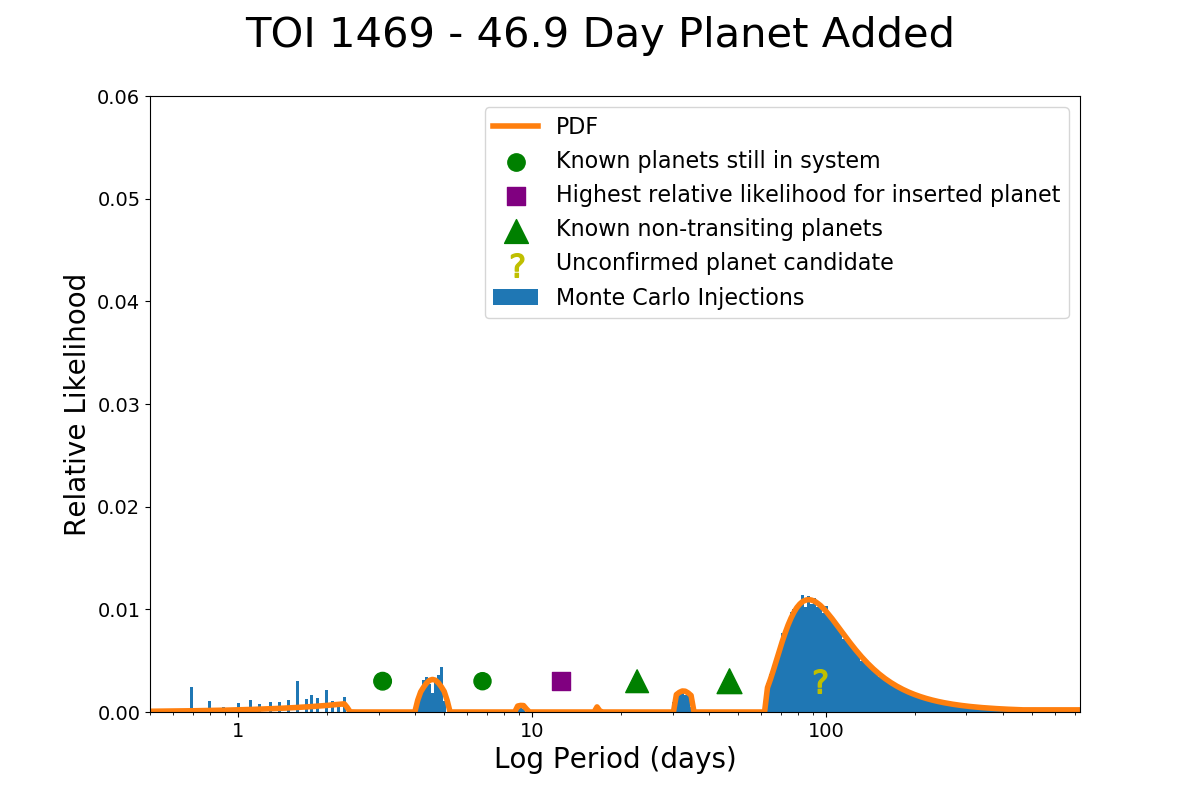}
    \caption{The TOI 1469 system (HD 219134) in various stages of \mname{} iterations.  Top Left: Running the \mname{} on the two transiting planets only, we find a large relative likelihood at 12.5 days, which does not correspond to any known planets.  Top Right: Adding a planet at that period, we find that \mname{} then finds a large relative likelihood very near the period of the inner non-transiting planet.  Bottom Left: Adding in that planet, \mname{} again shows a high relative likelihood at the location of the other confirmed non-transiting planet.  Bottom Right: After adding in all confirmed planets and the one inserted planet, \mname{} still shows that the 94-day period unconfirmed candidate would be the most likely insertion.}
    \label{fig:toi1469}
\end{figure*}

\mname{} is also able to provide analysis for planet candidates that are uncertain.  TOI 1469 was previously identified as a planetary system around the K3V star HD 219134; it contains 2 rocky super-Earth transiting planets that are analyzed in this sample, and at least 3 non-transiting massive planets exterior to these.  This system has been studied extensively as a result, and an unconfirmed planet candidate with its validity in doubt (as it was found in only one of these four studies of this system) was predicted with a period of $94.2 \pm 0.2$ days \citep{mot15, vog15, joh16, gil17b}.  Running \mname{} in an iterative way places injected planets most often at 12.5, 23, and then 45.5 days, adding in an injected planet every iteration; HD 219134 f and d have periods of $22.72 \pm 0.02$ and $46.86 \pm 0.03$ days, respectively.  In addition, \mname{} predicts a fourth injected planet at $\sim$ 87 days given the general period ratio sample.  These model runs are shown in Figure \ref{fig:toi1469}.

As a planet at 12.5 days actually minimizes the standard deviation of the period ratios in this system, if we then also insert a planet beyond HD 219134 f using the mean of the period ratios, as \citetalias{mul18} states the period ratios tend to be consistent within a system, then the inserted planet is most often found at a period of 93 days.  This inserted planet would only have a 5.1\% chance of transiting, given that the closer two planets do not transit.  The outermost known planet in the system has a period of $\sim$ 6 years so it would not be found by our analysis, as we only search out to 2 years due to the statistical constraints imposed by the current datasets and models.  Based on the planet radius and inclination analysis, the missing planet at 12.5 days would likely be in the smaller cluster (similar in size to the rocky super-Earths interior to it) and have a 75\% chance of transiting.  However, the transits of a 1.55-Earth-radius planet around a 0.78-solar-radius star would have a depth of $\sim$ 330 ppm and would likely only be visible from space \citep{gil17b}.

\subsection{Summary}

We tested the sensitivity and robustness of \mname{} by altering known multi-planet systems and seeing what was predicted when \mname{} was run on these systems.  We found that \mname{} responded well to the removal of one planet in the system, although the underlying statistical distributions were still noticeable if the original system wasn't regularly dynamically spaced.  Removing multiple consecutive planets caused a drop in the relative likelihood of injecting one planet across the large gap, but we showed that injecting two planets across that gap would recover the planets and near their correct periods as well.  The planet radius and inclination were mostly stable, although if one planet radius cluster consisting of two planets was removed from a system, then \mname{} would accordingly shift the relative likelihood away from finding planets of that size.  All three parameters were robust to changes on the order of a few percent.

\section{Applications to \textit{TESS} multi-planet systems} \label{sec:apps}

\subsection{Goals}

A compelling use of the predictive power of \mname{} as shown in Section \ref{sec:sav} is to analyze recently-discovered multi-planet systems from \textit{TESS} and find the systems with the highest relative likelihood of containing another detectable planet.  \textit{TESS} has a unique parameter space for exoplanet transit searches, in that it covers the entire sky in one-month sectors with magnitude limits of $m_T < 13$ for super-Earth transits and $m_T < 15$ magnitude for precise stellar photometry \citep{ric15}. This method can be used to predict the location of possible unseen planets in these systems when looking through the archival \textit{TESS} data for low-confidence signals which could be possible planets, as well as which systems are the best targets for follow-up ground-based observations to find new planets.  The \textit{TESS} extended mission will discover many more multi-planet systems that can be probed for these hidden planets.  

\subsection{Methodology}

To create our sample, we took the known multi-planet systems identified by \textit{TESS} as Objects of Interest (TOIs) located in the ExoFOP-TESS archive.  From the total list, we removed 4 systems containing planets or planet candidates with a physical radius larger than 5 $R_\oplus$, as these giant planets ($>5-6 R_\oplus$) tend to have a different distribution in period and multiplicity within a system \citep{ste12, don13}.  This left us with 52 systems containing at least two transiting planets or planet candidates with  planet radius less than 5 $R_\oplus$.

A majority of the \textit{TESS} systems have two currently known transiting planet candidates, with a few containing three and only two known systems with four transiting planet candidates.  Some systems (e.g., TOI 1469, as shown in Section \ref{subsec:savanv}) are known to have more non-transiting planets.  However, for the purpose of this analysis we used the method on only the transiting planets to gain inclination and planet radius distributions as well as the information on the likely orbital periods for the system.

We ran both orbital period models described in Section \ref{subsec:implement} to fully probe the period space for each system, along with the standard clustered planet radius distribution and Rayleigh inclination distribution with a fraction of isotropic systems.  In the period ratio model, if the largest peak in the period distribution interior to the outermost planet is comparable in relative likelihood to the exterior period peak after the last known planet, then we use the inner local maximum as the predicted period.  In the clustered period model, we find the specific period inside of one of the clusters that gives the maximum relative likelihood for a new planet.  For the planet radius distribution, we simply take the median of the distribution as the predicted planet radius, which is accurate for single clusters and tends to evenly weight the contributions from two clusters in the planet radius.

Given the predicted period, planet radius, and inclination distribution, we then calculate the probability that the planet would transit.  In addition, given the predicted planet radius and the host star's stellar radius, we calculate the most likely absolute transit depth.  We use a combination of these two factors to determine which systems should have the highest priority for follow-up, either from more thorough \textit{TESS} data searches or from ground-based observatories.

\section{Results} \label{sec:results}

\subsection{Period ratio model}

\begin{figure*}
    \centering
    \includegraphics[width=\textwidth]{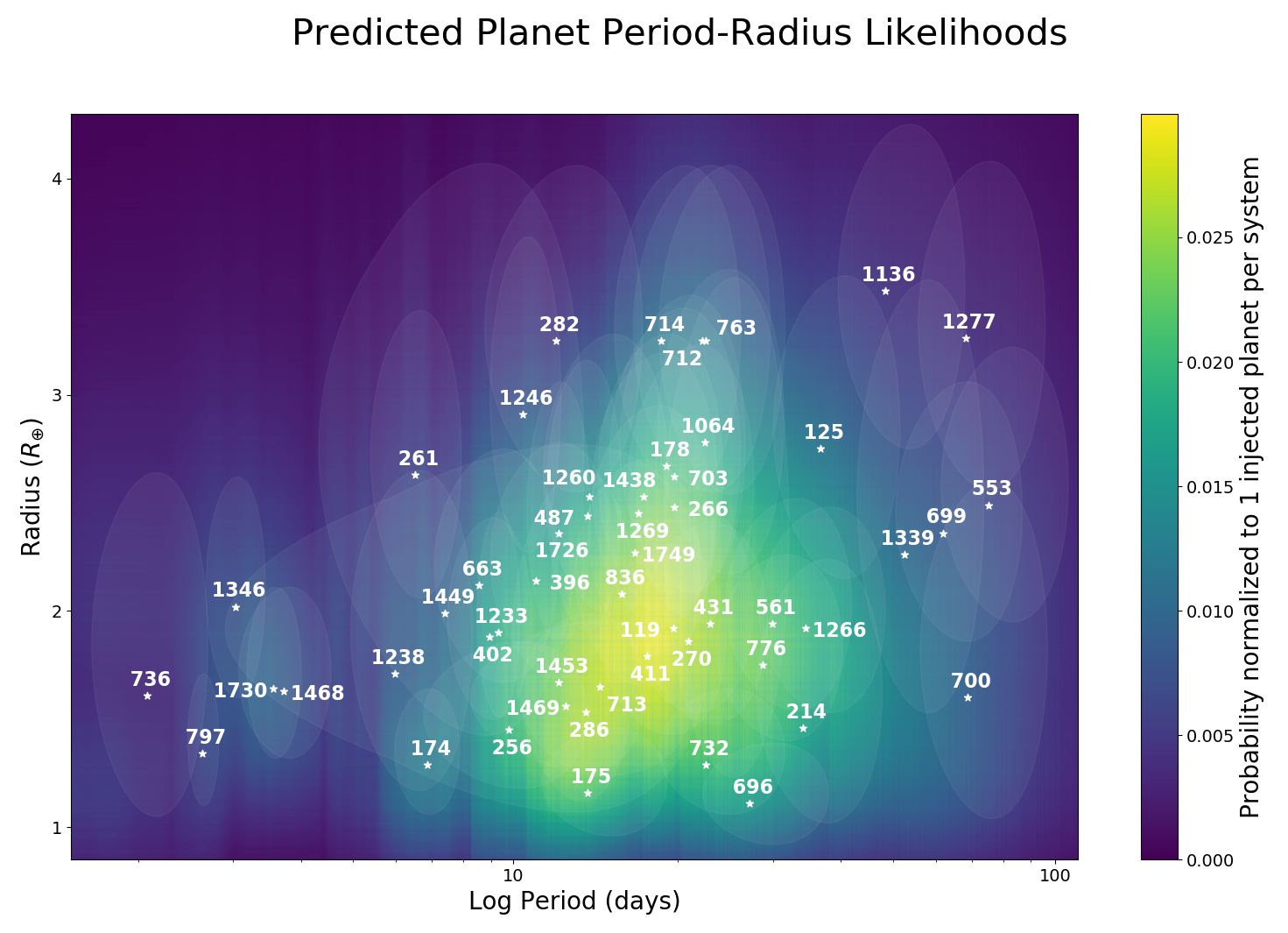}
    \caption{The two-dimensional probability space in log period and planet radius for the period ratio model, summed across all systems.  This shows a peak in the log period distribution between 10 and 40 days and in the radius distribution around 2 $R_\oplus$.  The individual system predictions are shown with white stars, with the ellipses representing $1\sigma$ uncertainties.}
    \label{fig:PRfig_epos}
\end{figure*}

\begin{figure*}[ht]
    \centering
    \includegraphics[width=0.95\textwidth]{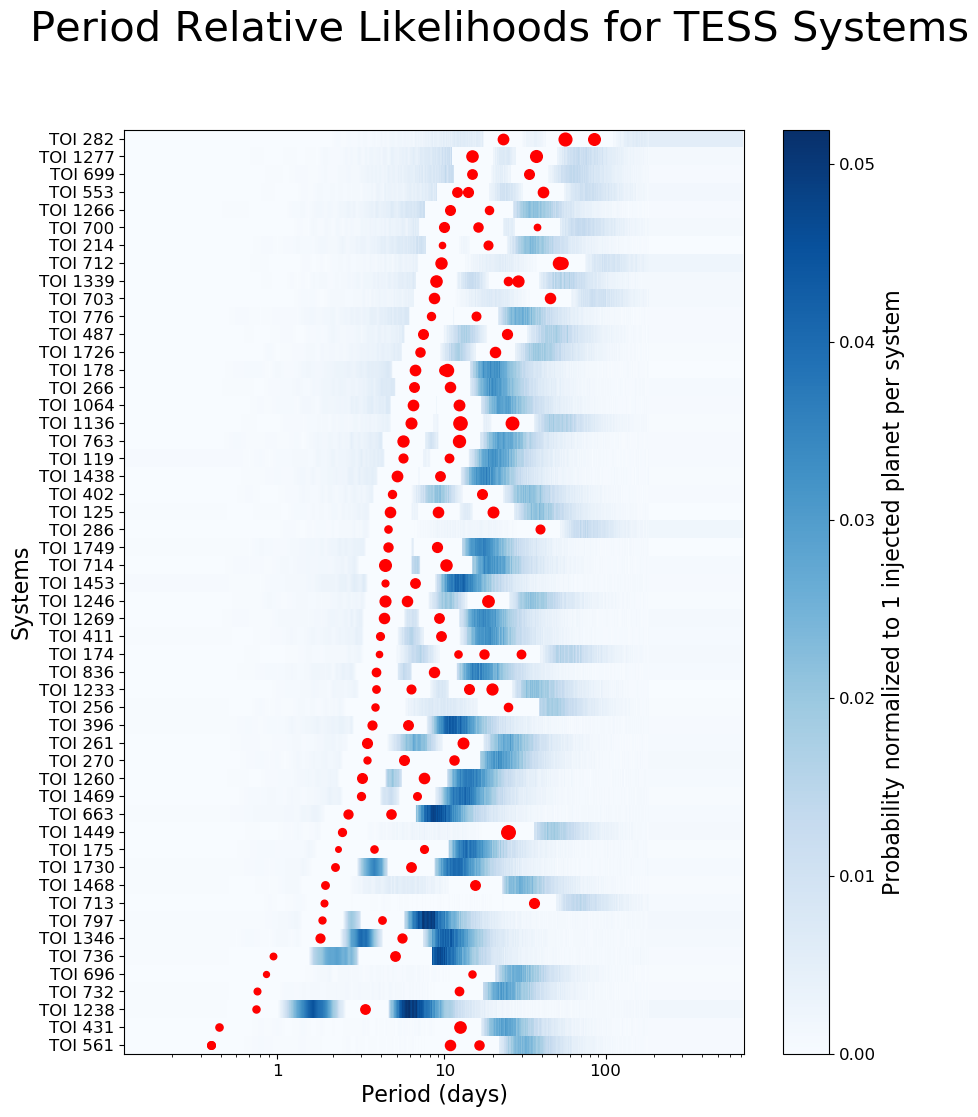}
    \caption{Probability of injecting a planet at a given period in log space using the period ratio model, normalized to 1 injection per system. Red dots indicate currently known planet or planet candidate periods, with relative marker size scaled to correspond to planet radius.  The planets are predicted most often to exist in period ratio gaps $\gtrsim$ 3, or exterior to the outermost known planet.  With long gaps in the period ratio, the likelihood of inserting just one planet in that gap using this model is low compared to an exterior insertion, but inserting multiple planets there is more likely.}
    \label{fig:logPfig}
\end{figure*}

\begin{figure*}[ht]
    \centering
    \includegraphics[width=0.95\textwidth]{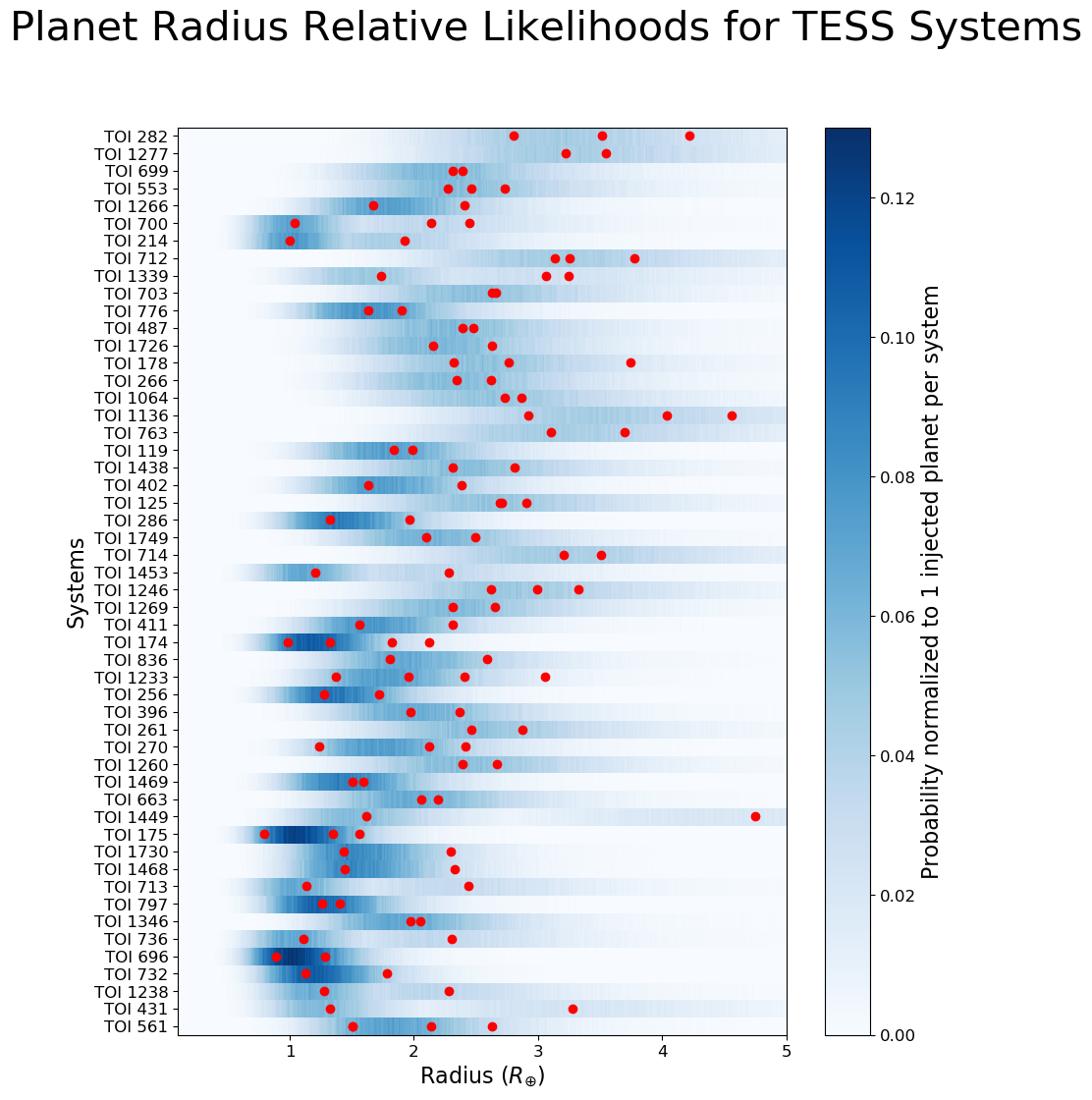}
    \caption{Probability of injecting a planet at a given planet radius across all systems, with red dots indicating currently known planet or planet candidate radii.  The systems with nearly equally sized planets show a smooth distribution, while systems with wide separation in planet radius show a double-peaked distribution.}
    \label{fig:Rfig}
\end{figure*}

\begin{figure*}
    \centering
    \includegraphics[width=0.95\textwidth]{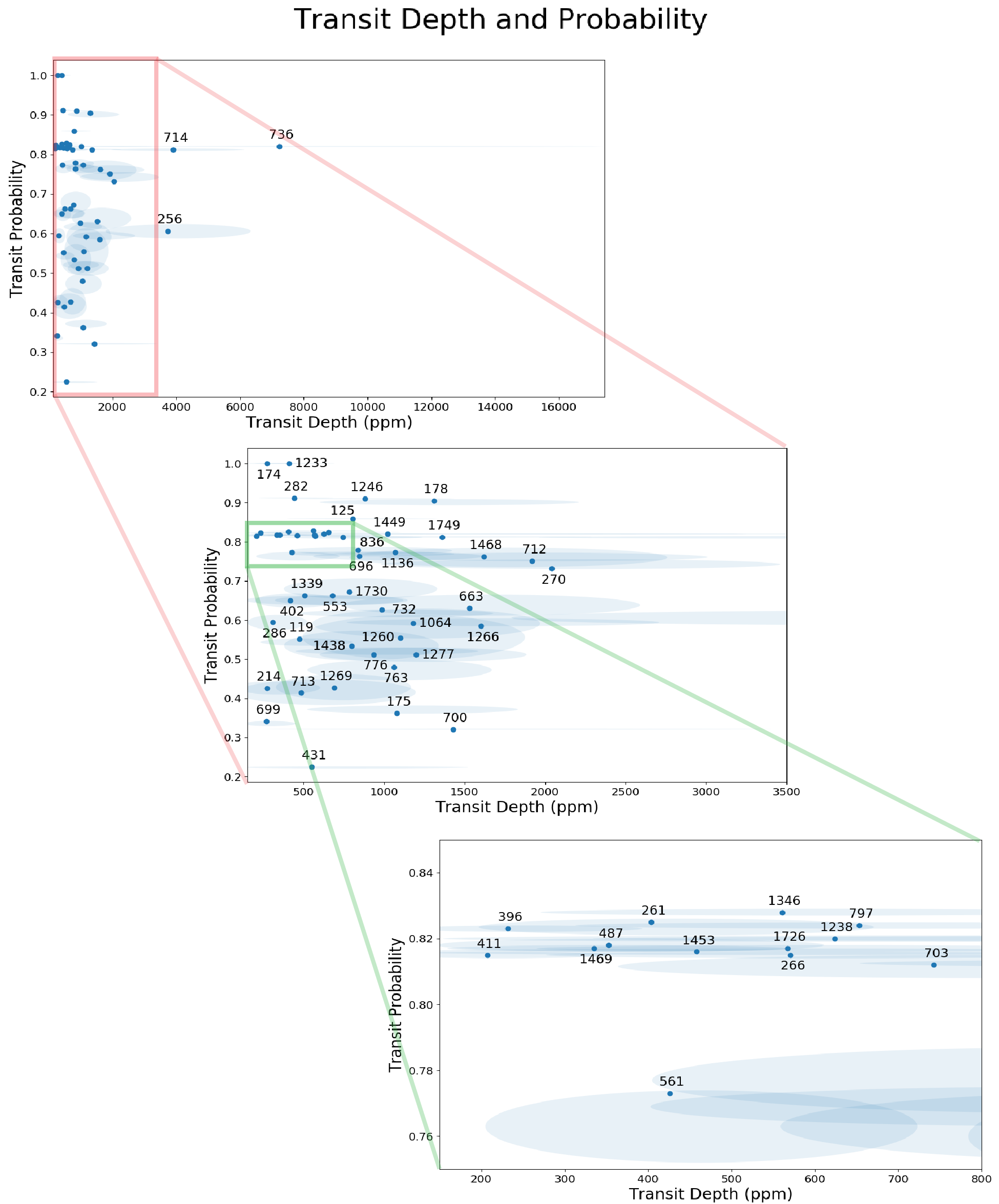}
    \caption{The predicted transit depths and probabilities for all systems using the period ratio model.  Systems with larger predicted depths and higher predicted probabilities are prioritized higher for follow-up observations.  Top: The three systems with transit depths greater than 3500 ppm.  Middle: A large majority of the systems have predicted transit depths between 250 and 2000 ppm.  Bottom: The cluster of systems with predicted transit depths between 200 and 800 ppm and predicted transit probabilities between 0.75 and 0.85.  Labels indicate the TOI number, with the ellipses representing $1\sigma$ uncertainties.}
    \label{fig:td_tp_epos}
\end{figure*}

Using the period ratio model from \citetalias{mul18}, we find that most systems have their peak in the period distribution between 10 and 40 days, with a relatively smooth log gradient in both directions away from the center of the peak.  The planet radius relative likelihood is high between 1-3 $R_\oplus$ and decreasing on both sides.  The full gradient across all systems in both dimensions for this model is shown in Figure \ref{fig:PRfig_epos}, with the predicted periods and planet radii for each system marked.  The full results, including 1$\sigma$ uncertainties calculated from the Monte Carlo iterations, are found in Table \ref{tab:vals_epos}.  The relative likelihood for each system in log period space calculated by the period ratio model is shown in Figure \ref{fig:logPfig}, as well as the relative likelihood in planet radius space (consistent between the different period models) in Figure \ref{fig:Rfig}.  The transit depth and transit probability for the period ratio model are shown in Figure \ref{fig:td_tp_epos}.

\begin{table*}[ht]
    {\centering
    \caption{Period Ratio Model Predictions}
    \label{tab:vals_epos}
    \tiny
    \begin{tabular}{|l|ccccc|}
        \hline
        System & Period (days) & Planet Radius ($R_\oplus$) & Stellar Radius ($R_\odot$) & Transit Depth (ppm) & Transit Probability\\
        \hline
        \object{TOI 561} & $30.0^{+12.1}_{-5.52}$ & $1.94^{+0.675}_{-0.503}$ & $0.856\pm0.014$ & $426^{+297}_{-221}$ & $0.773^{+0.001}_{-0.021}$ \\
		\object{TOI 431} & $23.0^{+9.28}_{-4.22}$ & $1.93^{+1.71}_{-0.808}$ & $0.751\pm0.015$ & $551^{+972}_{-461}$ & $0.225^{+0.001}_{-0.004}$ \\
		\object{TOI 1238} & $6.07^{+2.45}_{-1.12}$ & $1.71^{+1.04}_{-0.642}$ & $0.623\pm0.018$ & $624^{+761}_{-470}$ & $0.820^{+0.001}_{-0.001}$ \\
		\object{TOI 732} & $22.6^{+9.13}_{-4.15}$ & $1.29^{+0.449}_{-0.334}$ & $0.373\pm0.009$ & $987^{+691}_{-515}$ & $0.626^{+0.001}_{-0.019}$ \\
		\object{TOI 696} & $27.2^{+11.0}_{-5.00}$ & $1.10^{+0.384}_{-0.284}$ & $0.346\pm0.015$ & $844^{+592}_{-441}$ & $0.764^{+0.011}_{-0.001}$ \\
		\object{TOI 736} & $2.18^{+0.718}_{-0.54}$ & $1.61^{+1.14}_{-0.663}$ & $0.172\pm0.004$ & $7230^{+10200}_{-5980}$ & $0.820^{+0.001}_{-0.001}$ \\
		\object{TOI 1346} & $3.13^{+0.511}_{-0.469}$ & $2.01^{+0.704}_{-0.518}$ & $0.774\pm0.02$ & $561^{+394}_{-290}$ & $0.828^{+0.001}_{-0.001}$ \\
		\object{TOI 797} & $2.73^{+0.294}_{-0.266}$ & $1.34^{+0.463}_{-0.345}$ & $0.477\pm0.017$ & $653^{+454}_{-340}$ & $0.824^{+0.001}_{-0.001}$ \\
		\object{TOI 713} & $14.4^{+8.81}_{-11.5}$ & $1.66^{+1.22}_{-0.683}$ & $0.686\pm0.047$ & $484^{+713}_{-405}$ & $0.414^{+0.034}_{-0.031}$ \\
		\object{TOI 1468} & $3.82^{+0.934}_{-0.751}$ & $1.63^{+0.571}_{-0.418}$ & $0.369\pm0.011$ & $1620^{+1140}_{-836}$ & $0.762^{+0.023}_{-0.027}$ \\
		\object{TOI 1730} & $3.66^{+0.557}_{-0.484}$ & $1.64^{+0.566}_{-0.422}$ & $0.532\pm0.016$ & $785^{+545}_{-408}$ & $0.672^{+0.034}_{-0.019}$ \\
		\object{TOI 175} & $13.7^{+5.55}_{-2.52}$ & $1.15^{+0.400}_{-0.297}$ & $0.319\pm0.003$ & $1080^{+751}_{-558}$ & $0.362^{+0.02}_{-0.001}$ \\
		\object{TOI 1449} & $7.49^{+6.07}_{-3.17}$ & $1.99^{+2.18}_{-0.669}$ & $0.568\pm0.011$ & $1020^{+2230}_{-686}$ & $0.82^{+0.001}_{-0.001}$ \\
		\object{TOI 663} & $8.66^{+3.50}_{-1.59}$ & $2.11^{+0.730}_{-0.545}$ & $0.493\pm0.014$ & $1530^{+1060}_{-792}$ & $0.63^{+0.035}_{-0.019}$ \\
		\object{TOI 1469} & $12.5^{+5.04}_{-2.29}$ & $1.55^{+0.535}_{-0.400}$ & $0.772\pm0.041$ & $335^{+234}_{-176}$ & $0.817^{+0.001}_{-0.001}$ \\
		\object{TOI 1260} & $13.8^{+5.58}_{-2.54}$ & $2.53^{+0.851}_{-0.647}$ & $0.693\pm0.069$ & $1100^{+775}_{-606}$ & $0.555^{+0.058}_{-0.056}$ \\
		\object{TOI 270} & $21.0^{+8.47}_{-3.85}$ & $1.86^{+0.647}_{-0.480}$ & $0.375\pm0.01$ & $2040^{+1420}_{-1060}$ & $0.731^{+0.024}_{-0.001}$ \\
		\object{TOI 261} & $6.62^{+1.51}_{-1.23}$ & $2.62^{+0.866}_{-0.671}$ & $1.19\pm0.022$ & $404^{+267}_{-207}$ & $0.825^{+0.001}_{-0.004}$ \\
		\object{TOI 396} & $11.0^{+4.45}_{-2.02}$ & $2.16^{+0.747}_{-0.555}$ & $1.29\pm0.026$ & $232^{+161}_{-120}$ & $0.823^{+0.001}_{-0.001}$ \\
		\object{TOI 256} & $9.81^{+4.03}_{-3.05}$ & $1.45^{+0.504}_{-0.373}$ & $0.216\pm0.007$ & $3730^{+2610}_{-1940}$ & $0.606^{+0.018}_{-0.019}$ \\
		\object{TOI 1233} & $9.38^{+1.00}_{-0.906}$ & $1.89^{+0.659}_{-0.491}$ & $0.852\pm0.022$ & $410^{+286}_{-214}$ & $1$ \\
		\object{TOI 836} & $15.8^{+6.40}_{-2.91}$ & $2.08^{+0.721}_{-0.534}$ & $0.653\pm0.017$ & $839^{+584}_{-434}$ & $0.778^{+0.009}_{-0.011}$ \\
		\object{TOI 174} & $6.95^{+1.14}_{-0.983}$ & $1.29^{+0.446}_{-0.333}$ & $0.711\pm0.018$ & $272^{+189}_{-141}$ & $1$ \\
		\object{TOI 411} & $17.6^{+7.13}_{-3.24}$ & $1.79^{+0.618}_{-0.461}$ & $1.13\pm0.049$ & $207^{+145}_{-108}$ & $0.815^{+0.001}_{-0.001}$ \\
		\object{TOI 1269} & $17.0^{+6.88}_{-3.13}$ & $2.45^{+0.833}_{-0.628}$ & $0.849\pm0.045$ & $691^{+476}_{-362}$ & $0.427^{+0.035}_{-0.032}$ \\
		\object{TOI 1246} & $10.4^{+1.93}_{-1.43}$ & $2.91^{+0.921}_{-0.737}$ & $0.893\pm0.028$ & $880^{+560}_{-449}$ & $0.91^{+0.001}_{-0.001}$ \\
		\object{TOI 1453} & $12.1^{+4.91}_{-2.23}$ & $1.67^{+1.08}_{-0.655}$ & $0.71\pm0.048$ & $458^{+595}_{-365}$ & $0.816^{+0.001}_{-0.001}$ \\
		\object{TOI 714} & $18.7^{+7.58}_{-3.45}$ & $3.25^{+0.915}_{-0.802}$ & $0.473\pm0.014$ & $3910^{+2220}_{-1950}$ & $0.811^{+0.003}_{-0.004}$ \\
		\object{TOI 1749} & $16.7^{+6.74}_{-3.06}$ & $2.27^{+0.775}_{-0.586}$ & $0.561\pm0.017$ & $1360^{+931}_{-706}$ & $0.812^{+0.002}_{-0.001}$ \\
		\object{TOI 286} & $13.6^{+7.94}_{-5.35}$ & $1.52^{+0.532}_{-0.391}$ & $0.789\pm0.011$ & $309^{+216}_{-159}$ & $0.594^{+0.019}_{-0.019}$ \\
		\object{TOI 125} & $36.8^{+14.9}_{-6.77}$ & $2.75^{+0.901}_{-0.702}$ & $0.883\pm0.002$ & $804^{+527}_{-411}$ & $0.859^{+0.001}_{-0.001}$ \\
		\object{TOI 402} & $9.04^{+1.87}_{-1.55}$ & $1.88^{+0.649}_{-0.487}$ & $0.840\pm0.013$ & $415^{+287}_{-216}$ & $0.65^{+0.018}_{-0.019}$ \\
		\object{TOI 1438} & $17.4^{+7.02}_{-3.19}$ & $2.53^{+0.849}_{-0.648}$ & $0.816\pm0.046$ & $797^{+543}_{-418}$ & $0.534^{+0.039}_{-0.038}$ \\
		\object{TOI 119} & $19.7^{+7.96}_{-3.62}$ & $1.91^{+0.664}_{-0.490}$ & $0.800\pm0.011$ & $475^{+330}_{-244}$ & $0.552^{+0.001}_{-0.018}$ \\
		\object{TOI 763} & $22.6^{+9.14}_{-4.16}$ & $3.24^{+0.916}_{-0.802}$ & $0.907\pm0.047$ & $1060^{+609}_{-536}$ & $0.48^{+0.019}_{-0.034}$ \\
		\object{TOI 1136} & $48.5^{+19.6}_{-8.91}$ & $3.47^{+0.878}_{-0.829}$ & $0.968\pm0.017$ & $1070^{+541}_{-511}$ & $0.773^{+0.001}_{-0.021}$ \\
		\object{TOI 1064} & $22.5^{+9.11}_{-4.14}$ & $2.77^{+0.906}_{-0.709}$ & $0.737\pm0.057$ & $1180^{+789}_{-628}$ & $0.592^{+0.037}_{-0.039}$ \\
		\object{TOI 266} & $19.8^{+8.00}_{-3.64}$ & $2.48^{+0.848}_{-0.635}$ & $0.944\pm0.016$ & $571^{+392}_{-293}$ & $0.815^{+0.001}_{-0.001}$ \\
		\object{TOI 178} & $19.1^{+7.71}_{-3.51}$ & $2.67^{+0.885}_{-0.681}$ & $0.672\pm0.061$ & $1310^{+901}_{-710}$ & $0.905^{+0.004}_{-0.012}$ \\
		\object{TOI 1726} & $12.1^{+1.84}_{-1.60}$ & $2.36^{+0.801}_{-0.610}$ & $0.903\pm0.055$ & $567^{+391}_{-301}$ & $0.817^{+0.001}_{-0.001}$ \\
		\object{TOI 487} & $13.7^{+2.23}_{-2.30}$ & $2.43^{+0.823}_{-0.623}$ & $1.18\pm0.165$ & $353^{+258}_{-206}$ & $0.818^{+0.002}_{-0.002}$ \\
		\object{TOI 776} & $28.8^{+11.7}_{-5.30}$ & $1.77^{+0.611}_{-0.458}$ & $0.526\pm0.012$ & $936^{+649}_{-487}$ & $0.512^{+0.018}_{-0.001}$ \\
		\object{TOI 703} & $19.8^{+6.44}_{-4.86}$ & $2.64^{+0.880}_{-0.672}$ & $0.883\pm0.019$ & $743^{+496}_{-379}$ & $0.812^{+0.003}_{-0.004}$ \\
		\object{TOI 1339} & $52.6^{+21.3}_{-9.68}$ & $2.26^{+1.37}_{-0.830}$ & $0.914\pm0.025$ & $506^{+615}_{-373}$ & $0.662^{+0.001}_{-0.024}$ \\
		\object{TOI 712} & $22.3^{+7.51}_{-5.77}$ & $3.24^{+0.916}_{-0.799}$ & $0.674\pm0.011$ & $1920^{+1090}_{-949}$ & $0.751^{+0.021}_{-0.001}$ \\
		\object{TOI 214} & $34.2^{+13.8}_{-6.28}$ & $1.46^{+0.879}_{-0.542}$ & $0.802\pm0.026$ & $273^{+331}_{-204}$ & $0.426^{+0.018}_{-0.016}$ \\
		\object{TOI 700} & $68.9^{+27.9}_{-12.7}$ & $1.59^{+1.10}_{-0.661}$ & $0.384\pm0.009$ & $1430^{+1980}_{-1190}$ & $0.321^{+0.001}_{-0.001}$ \\
		\object{TOI 1266} & $34.6^{+14.0}_{-6.37}$ & $1.92^{+0.661}_{-0.493}$ & $0.436\pm0.013$ & $1600^{+1110}_{-831}$ & $0.585^{+0.019}_{-0.001}$ \\
		\object{TOI 553} & $75.3^{+30.5}_{-13.9}$ & $2.48^{+0.838}_{-0.639}$ & $0.866\pm0.047$ & $681^{+466}_{-359}$ & $0.662^{+0.001}_{-0.024}$ \\
		\object{TOI 699} & $62.0^{+25.0}_{-11.4}$ & $2.36^{+0.802}_{-0.604}$ & $1.31\pm0.027$ & $268^{+183}_{-138}$ & $0.341^{+0.001}_{-0.012}$ \\
		\object{TOI 1277} & $68.3^{+27.6}_{-12.6}$ & $3.26^{+0.917}_{-0.800}$ & $0.858\pm0.043$ & $1200^{+685}_{-600}$ & $0.512^{+0.018}_{-0.019}$ \\
		\object{TOI 282} & $12.0^{+5.38}_{-3.23}$ & $3.25^{+0.913}_{-0.805}$ & $1.41\pm0.032$ & $440^{+248}_{-219}$ & $0.912^{+0.001}_{-0.001}$ \\
        \hline
    \end{tabular}
    }
    \\[10pt]
    \small
    \textbf{Notes}: The predicted peaks in the period and planet radius distribution from the period ratio model, along with the calculated transit depths and transit probabilities from those predictions.  1$\sigma$ uncertainties were calculated from the Monte Carlo iterations.  Stellar radii and their uncertainties were taken from the ExoFOP-TESS archive.  Systems with transit probability of 1 are 4-planet systems with mutual inclination distributions modeled solely as a Rayleigh distribution (i.e., no isotropic fraction), and have geometric probabilities of 1 at the current period prediction amongst the entire inclination distribution.
\end{table*}

Using this model, we find for each system that the relative likelihood of finding a planet at a given period tends to peak at the center of a gap between two known planets, especially if the period ratio between those two is $\sim 4$, as well as a factor of $\sim 2$ exterior to the outermost known planet, as shown in Figure \ref{fig:logPfig}.  The likelihood measured vs separation in mutual Hill radii peaks at 20, similar to the analysis by \citet{gil20}.  For large period ratios between planets, the normalization of the Monte Carlo iterations to 1 injected planet smooths out the distribution and significantly lowers the local maximum in the relative likelihood.  The planet radius relative likelihoods show a relatively wider spread in the distribution between $\sim 1-5 R_\oplus$.

\subsection{Clustered periods model}

\begin{figure*}[ht]
    \centering
    \includegraphics[width=\textwidth]{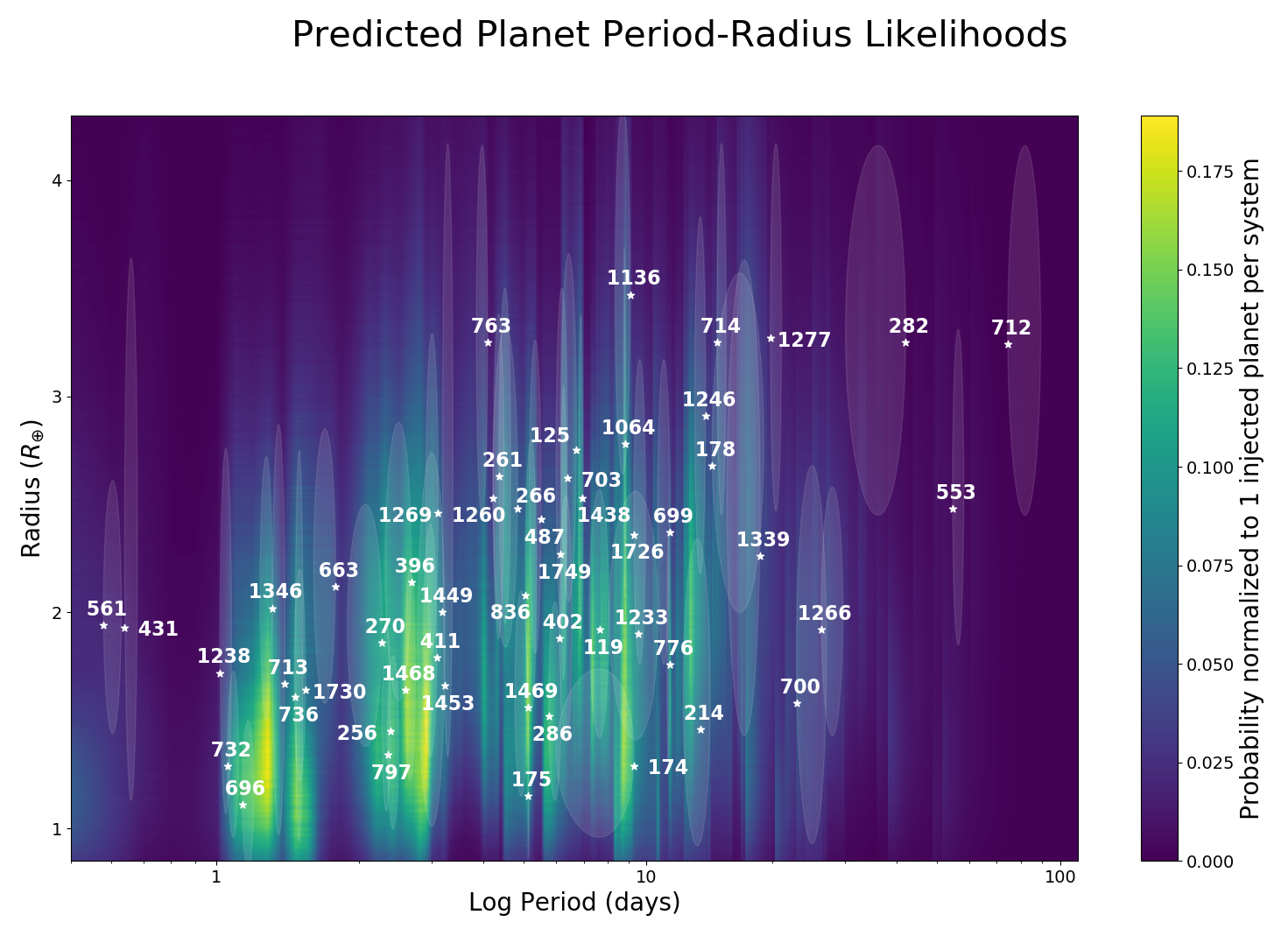}
    \caption{The two-dimensional probability space in log period and planet radius for the clustered periods model, summed across all systems.  The peak in the log period distribution occurs between 1 and 20 days, shorter than the period ratio model peak of 10-40 days.  The individual system predictions are shown with white stars, with the ellipses representing $1\sigma$ uncertainties.}
    \label{fig:PRfig_hfr}
\end{figure*}

\begin{figure*}[ht]
    \centering
    \includegraphics[width=0.95\textwidth]{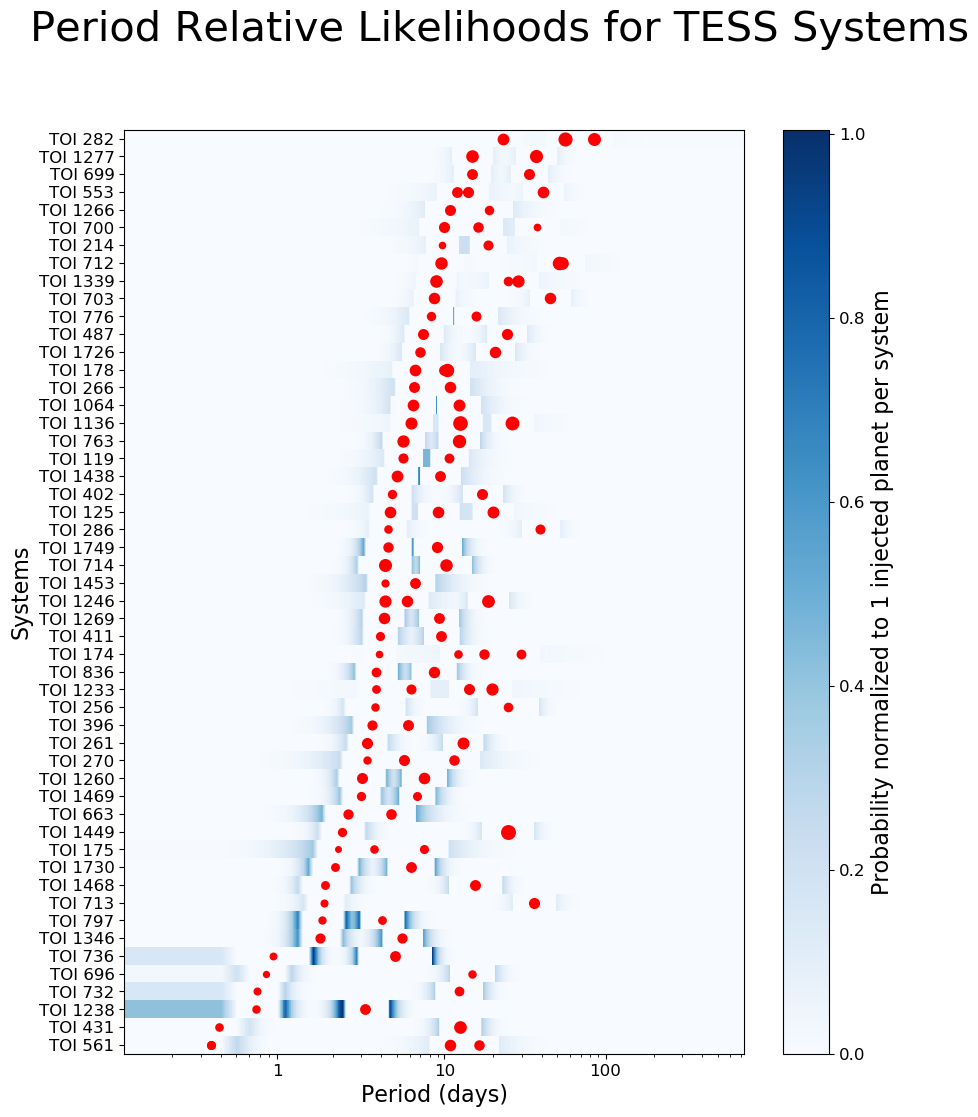}
    \caption{Probability of injecting a planet at a given period in log space using the clustered periods model, normalized to 1 injection per system. Red dots indicate currently known planet or planet candidate periods, with relative marker size scaled to correspond to planet radius.  Planets are most often predicted just outside the dynamical stability inner limit for each planet, as the clusters here are defined with respect to the known planets in the system.}
    \label{fig:logPfig_hfr}
\end{figure*}

\begin{figure*}
    \centering
    \includegraphics[width=0.95\textwidth]{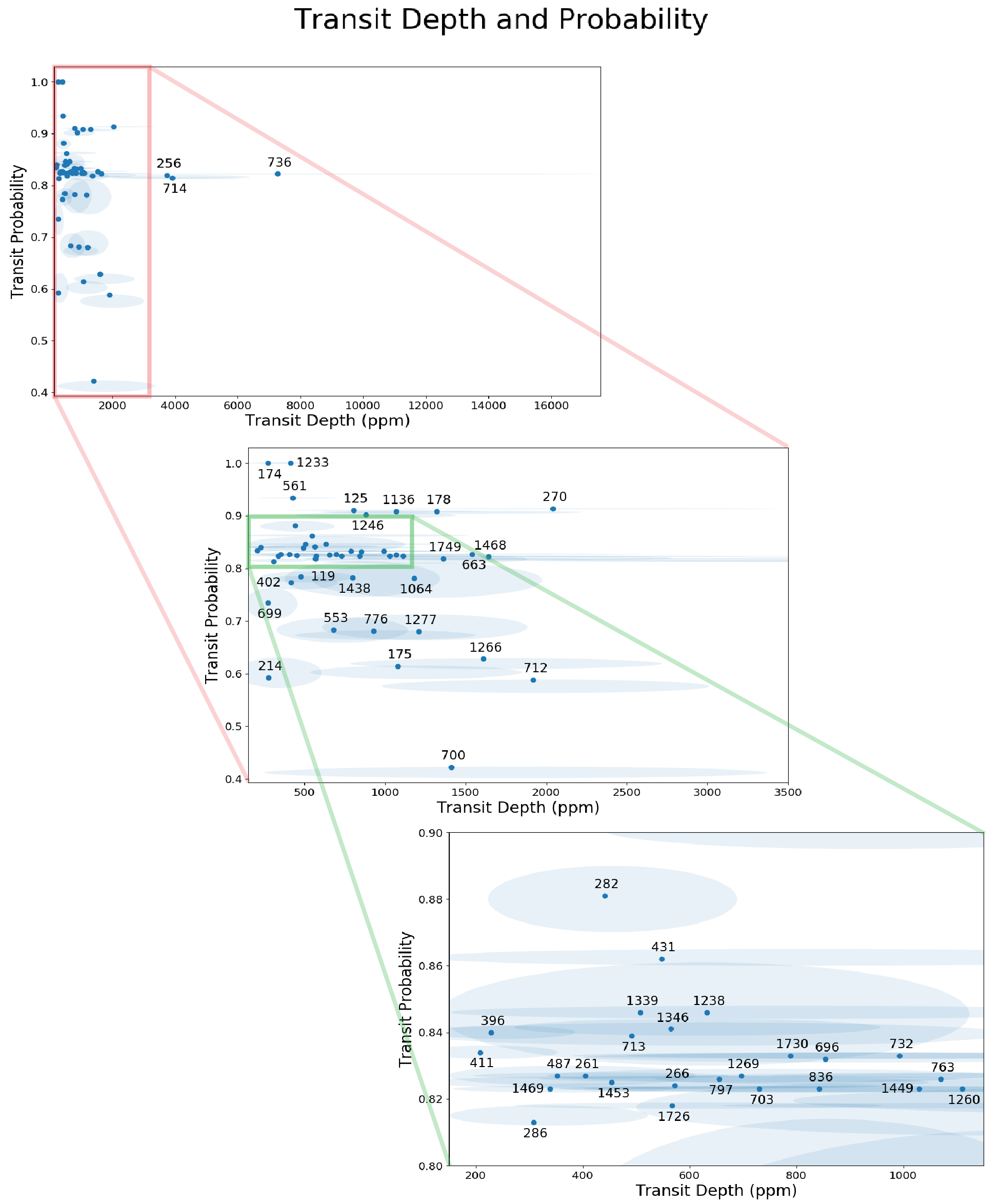}
    \caption{The predicted transit depths and probabilities for all systems using the clustered periods model.  Systems with larger predicted depths and higher predicted probabilities are prioritized higher for follow-up observations.  Top: The three systems with transit depths greater than 3500 ppm.  Middle: A large majority of the systems have predicted transit depths between 250 and 2000 ppm.  Bottom: The cluster of systems with predicted transit depths between 200 and 1250 ppm and predicted transit probabilities between 0.8 and 0.9.  Labels indicate the TOI number, with the ellipses representing $1\sigma$ uncertainties.}
    \label{fig:td_tp_hfr}
\end{figure*}

The clustered periods model from \citetalias{hef19} tends to predict planets closer to their host stars, with peaks between 1 and 20 days (see Figure \ref{fig:PRfig_hfr}).  As a majority of these systems have a planet within 10 days, this model prefers to place another planet within that cluster.  Due to the varied positions of the second planet in the system, the clustered periods model has non-negligible probabilities of forming a second narrow cluster or combining the two planets into one cluster, which shows through in the period predictions.  The log period and planet radius parameter space for this model is shown in Figure \ref{fig:PRfig_hfr}), with the predicted periods and planet radii for each system marked.  The full results, including 1$\sigma$ uncertainties calculated from the Monte Carlo iterations, are found in Table \ref{tab:vals_hfr}.  The relative likelihood for each system in log period space calculated by the clustered periods model is shown in Figure \ref{fig:logPfig_hfr}, and the transit depth and transit probability for this model are shown in Figure \ref{fig:td_tp_hfr}.

\begin{table*}[ht]
    {\centering
    \caption{Predictions from Clustered Periods Model}
    \label{tab:vals_hfr}
    \tiny
    \begin{tabular}{|l|ccccc|}
        \hline
        System & Period (days) & Planet Radius ($R_\oplus$) & Stellar Radius ($R_\odot$) & Transit Depth (ppm) & Transit Probability\\
        \hline
        \object{TOI 561} & $0.587^{+0.051}_{-0.001}$ & $1.94^{+0.672}_{-0.500}$ & $0.856\pm0.014$ & $425^{+295}_{-220}$ & $0.934^{+0.001}_{-0.001}$ \\
		\object{TOI 431} & $0.644^{+0.042}_{-0.001}$ & $1.93^{+1.71}_{-0.799}$ & $0.751\pm0.015$ & $545^{+972}_{-453}$ & $0.862^{+0.003}_{-0.002}$ \\
		\object{TOI 1238} & $1.02^{+0.067}_{-0.001}$ & $1.72^{+1.04}_{-0.651}$ & $0.623\pm0.018$ & $630^{+765}_{-479}$ & $0.846^{+0.002}_{-0.002}$ \\
		\object{TOI 732} & $1.06^{+0.063}_{-0.001}$ & $1.28^{+0.447}_{-0.331}$ & $0.373\pm0.009$ & $982^{+686}_{-509}$ & $0.833^{+0.001}_{-0.001}$ \\
		\object{TOI 696} & $1.15^{+0.076}_{-0.001}$ & $1.10^{+0.389}_{-0.285}$ & $0.346\pm0.015$ & $845^{+600}_{-442}$ & $0.832^{+0.002}_{-0.001}$ \\
		\object{TOI 736} & $1.50^{+0.066}_{-0.001}$ & $1.60^{+1.14}_{-0.663}$ & $0.172\pm0.004$ & $7200^{+10200}_{-5970}$ & $0.822^{+0.001}_{-0.001}$ \\
		\object{TOI 1346} & $1.33^{+0.001}_{-0.088}$ & $2.01^{+0.700}_{-0.519}$ & $0.774\pm0.020$ & $560^{+391}_{-291}$ & $0.841^{+0.002}_{-0.002}$ \\
		\object{TOI 797} & $2.44^{+0.158}_{-0.001}$ & $1.34^{+0.463}_{-0.345}$ & $0.477\pm0.017$ & $651^{+454}_{-340}$ & $0.826^{+0.001}_{-0.001}$ \\
		\object{TOI 713} & $1.42^{+0.001}_{-0.097}$ & $1.67^{+1.19}_{-0.701}$ & $0.686\pm0.047$ & $494^{+707}_{-420}$ & $0.839^{+0.004}_{-0.003}$ \\
		\object{TOI 1468} & $2.68^{+0.154}_{-0.001}$ & $1.63^{+0.568}_{-0.421}$ & $0.369\pm0.011$ & $1630^{+1140}_{-844}$ & $0.822^{+0.002}_{-0.007}$ \\
		\object{TOI 1730} & $1.58^{+0.001}_{-0.096}$ & $1.63^{+0.567}_{-0.421}$ & $0.532\pm0.016$ & $782^{+545}_{-406}$ & $0.833^{+0.001}_{-0.001}$ \\
		\object{TOI 175} & $5.22^{+0.031}_{-0.001}$ & $1.15^{+0.401}_{-0.297}$ & $0.319\pm0.003$ & $1080^{+753}_{-558}$ & $0.614^{+0.001}_{-0.024}$ \\
		\object{TOI 1449} & $3.28^{+0.199}_{-0.001}$ & $2.00^{+2.17}_{-0.673}$ & $0.568\pm0.011$ & $1030^{+2230}_{-694}$ & $0.823^{+0.001}_{-0.001}$ \\
		\object{TOI 663} & $1.85^{+0.001}_{-0.204}$ & $2.12^{+0.733}_{-0.543}$ & $0.493\pm0.014$ & $1530^{+1060}_{-789}$ & $0.827^{+0.001}_{-0.001}$ \\
		\object{TOI 1469} & $5.21^{+0.001}_{-0.350}$ & $1.55^{+0.543}_{-0.401}$ & $0.772\pm0.041$ & $335^{+237}_{-177}$ & $0.823^{+0.001}_{-0.001}$ \\
		\object{TOI 1260} & $4.32^{+0.264}_{-0.001}$ & $2.53^{+0.852}_{-0.651}$ & $0.693\pm0.069$ & $1100^{+776}_{-609}$ & $0.823^{+0.003}_{-0.014}$ \\
		\object{TOI 270} & $2.37^{+0.001}_{-0.405}$ & $1.86^{+0.643}_{-0.476}$ & $0.375\pm0.010$ & $2030^{+1410}_{-1050}$ & $0.913^{+0.001}_{-0.001}$ \\
		\object{TOI 261} & $4.46^{+0.295}_{-0.001}$ & $2.62^{+0.874}_{-0.672}$ & $1.19\pm0.022$ & $403^{+269}_{-207}$ & $0.827^{+0.001}_{-0.001}$ \\
		\object{TOI 396} & $2.77^{+0.001}_{-0.360}$ & $2.16^{+0.746}_{-0.557}$ & $1.29\pm0.026$ & $232^{+161}_{-120}$ & $0.840^{+0.002}_{-0.002}$ \\
		\object{TOI 256} & $2.48^{+0.001}_{-0.113}$ & $1.45^{+0.502}_{-0.374}$ & $0.216\pm0.007$ & $3740^{+2600}_{-1940}$ & $0.819^{+0.001}_{-0.007}$ \\
		\object{TOI 1233} & $9.58^{+1.11}_{-1.43}$ & $1.90^{+0.658}_{-0.488}$ & $0.852\pm0.022$ & $411^{+286}_{-213}$ & $1$ \\
		\object{TOI 836} & $5.15^{+0.329}_{-0.001}$ & $2.07^{+0.722}_{-0.532}$ & $0.653\pm0.017$ & $835^{+584}_{-431}$ & $0.823^{+0.001}_{-0.001}$ \\
		\object{TOI 174} & $3.12^{+0.001}_{-0.223}$ & $1.53^{+1.00}_{-0.602}$ & $0.711\pm0.018$ & $383^{+503}_{-303}$ & $1$ \\
		\object{TOI 411} & $3.18^{+0.001}_{-0.224}$ & $1.79^{+0.624}_{-0.460}$ & $1.13\pm0.049$ & $207^{+146}_{-108}$ & $0.833^{+0.002}_{-0.002}$ \\
		\object{TOI 1269} & $3.20^{+0.001}_{-0.200}$ & $2.46^{+0.828}_{-0.631}$ & $0.849\pm0.045$ & $695^{+474}_{-365}$ & $0.83^{+0.002}_{-0.002}$ \\
		\object{TOI 1246} & $13.9^{+0.001}_{-0.827}$ & $2.90^{+0.919}_{-0.733}$ & $0.893\pm0.028$ & $874^{+557}_{-446}$ & $0.900^{+0.005}_{-0.007}$ \\
		\object{TOI 1453} & $3.32^{+0.001}_{-0.450}$ & $1.67^{+1.08}_{-0.654}$ & $0.710\pm0.048$ & $457^{+593}_{-364}$ & $0.827^{+0.002}_{-0.002}$ \\
		\object{TOI 714} & $14.8^{+0.766}_{-0.001}$ & $3.24^{+0.921}_{-0.795}$ & $0.473\pm0.014$ & $3890^{+2230}_{-1920}$ & $0.814^{+0.001}_{-0.001}$ \\
		\object{TOI 1749} & $6.25^{+0.198}_{-0.001}$ & $2.27^{+0.782}_{-0.582}$ & $0.561\pm0.017$ & $1360^{+939}_{-701}$ & $0.818^{+0.001}_{-0.001}$ \\
		\object{TOI 286} & $5.85^{+0.415}_{-0.001}$ & $1.52^{+0.529}_{-0.391}$ & $0.789\pm0.011$ & $308^{+215}_{-159}$ & $0.813^{+0.005}_{-0.001}$ \\
		\object{TOI 125} & $6.81^{+0.001}_{-0.560}$ & $2.74^{+0.912}_{-0.699}$ & $0.883\pm0.002$ & $800^{+532}_{-408}$ & $0.910^{+0.001}_{-0.001}$ \\
		\object{TOI 402} & $6.21^{+0.443}_{-0.001}$ & $1.88^{+0.654}_{-0.485}$ & $0.840\pm0.013$ & $415^{+289}_{-215}$ & $0.773^{+0.012}_{-0.013}$ \\
		\object{TOI 1438} & $7.04^{+0.019}_{-0.149}$ & $2.53^{+0.849}_{-0.652}$ & $0.816\pm0.046$ & $796^{+542}_{-420}$ & $0.782^{+0.026}_{-0.036}$ \\
		\object{TOI 119} & $7.73^{+0.412}_{-0.426}$ & $1.91^{+0.655}_{-0.495}$ & $0.800\pm0.011$ & $475^{+326}_{-246}$ & $0.784^{+0.010}_{-0.011}$ \\
		\object{TOI 763} & $4.19^{+0.001}_{-0.257}$ & $3.25^{+0.908}_{-0.800}$ & $0.907\pm0.047$ & $1070^{+606}_{-536}$ & $0.826^{+0.002}_{-0.002}$ \\
		\object{TOI 1136} & $9.15^{+0.001}_{-0.722}$ & $3.47^{+0.886}_{-0.831}$ & $0.968\pm0.017$ & $1060^{+545}_{-512}$ & $0.910^{+0.001}_{-0.001}$ \\
		\object{TOI 1064} & $8.91^{+0.001}_{-0.078}$ & $2.78^{+0.908}_{-0.707}$ & $0.737\pm0.057$ & $1180^{+792}_{-627}$ & $0.782^{+0.030}_{-0.037}$ \\
		\object{TOI 266} & $4.93^{+0.001}_{-0.624}$ & $2.47^{+0.846}_{-0.635}$ & $0.944\pm0.016$ & $569^{+390}_{-293}$ & $0.824^{+0.001}_{-0.001}$ \\
		\object{TOI 178} & $14.4^{+4.78}_{-0.001}$ & $2.67^{+0.889}_{-0.680}$ & $0.672\pm0.061$ & $1310^{+905}_{-709}$ & $0.908^{+0.001}_{-0.004}$ \\
		\object{TOI 1726} & $9.34^{+0.634}_{-0.001}$ & $2.36^{+0.808}_{-0.605}$ & $0.903\pm0.055$ & $566^{+394}_{-299}$ & $0.818^{+0.001}_{-0.001}$ \\
		\object{TOI 487} & $5.61^{+0.001}_{-0.353}$ & $2.43^{+0.828}_{-0.624}$ & $1.18\pm0.165$ & $351^{+259}_{-205}$ & $0.827^{+0.003}_{-0.003}$ \\
		\object{TOI 776} & $11.4^{+0.001}_{-0.194}$ & $1.77^{+0.614}_{-0.458}$ & $0.526\pm0.012$ & $940^{+653}_{-488}$ & $0.681^{+0.001}_{-0.017}$ \\
		\object{TOI 703} & $6.48^{+0.001}_{-0.395}$ & $2.65^{+0.884}_{-0.677}$ & $0.883\pm0.019$ & $745^{+499}_{-383}$ & $0.821^{+0.001}_{-0.001}$ \\
		\object{TOI 1339} & $18.8^{+0.001}_{-3.16}$ & $2.26^{+1.37}_{-0.830}$ & $0.914\pm0.025$ & $505^{+616}_{-373}$ & $0.846^{+0.015}_{-0.016}$ \\
		\object{TOI 712} & $74.5^{+15.0}_{-0.001}$ & $3.24^{+0.920}_{-0.794}$ & $0.674\pm0.011$ & $1910^{+1090}_{-941}$ & $0.588^{+0.001}_{-0.025}$ \\
		\object{TOI 214} & $13.5^{+0.767}_{-1.23}$ & $1.46^{+0.874}_{-0.545}$ & $0.802\pm0.026$ & $275^{+330}_{-206}$ & $0.592^{+0.038}_{-0.019}$ \\
		\object{TOI 700} & $23.0^{+4.17}_{-0.001}$ & $1.57^{+1.11}_{-0.642}$ & $0.384\pm0.009$ & $1390^{+1960}_{-1140}$ & $0.421^{+0.001}_{-0.021}$ \\
		\object{TOI 1266} & $26.4^{+3.35}_{-0.001}$ & $1.92^{+0.665}_{-0.491}$ & $0.436\pm0.013$ & $1600^{+1120}_{-827}$ & $0.629^{+0.001}_{-0.019}$ \\
		\object{TOI 553} & $54.8^{+3.52}_{-0.001}$ & $2.48^{+0.835}_{-0.634}$ & $0.866\pm0.047$ & $678^{+463}_{-355}$ & $0.683^{+0.024}_{-0.024}$ \\
		\object{TOI 699} & $11.4^{+0.001}_{-0.743}$ & $2.36^{+0.801}_{-0.607}$ & $1.31\pm0.027$ & $270^{+183}_{-139}$ & $0.735^{+0.028}_{-0.032}$ \\
		\object{TOI 1277} & $19.9^{+1.28}_{-0.001}$ & $3.26^{+0.909}_{-0.797}$ & $0.858\pm0.043$ & $1200^{+679}_{-599}$ & $0.680^{+0.033}_{-0.017}$ \\
		\object{TOI 282} & $42.2^{+0.001}_{-12.0}$ & $3.24^{+0.915}_{-0.798}$ & $1.41\pm0.032$ & $439^{+249}_{-217}$ & $0.881^{+0.009}_{-0.011}$ \\
		\hline
	\end{tabular}
	}
	\\[10pt]
	\small
    \textbf{Notes}: The predicted peaks in the period and planet radius distribution from the clustered periods model, along with the calculated transit depths and transit probabilities from those predictions.  1$\sigma$ uncertainties were calculated from the Monte Carlo iterations.  Stellar radii and their uncertainties were taken from the ExoFOP-TESS archive.  Systems with transit probability of 1 are 4-planet systems with mutual inclination distributions modeled solely as a Rayleigh distribution (i.e., no isotropic fraction), and have geometric probabilities of 1 at the current period prediction amongst the entire inclination distribution.
\end{table*}

For this model, we find that the predicted peak in the planet period distribution tends to be at or near the dynamical stability limit for one of the planets in the system (see Figure \ref{fig:logPfig_hfr}), and the likelihood measured vs separation in mutual Hill radii peaks at the imposed inner limit of 8.  For most cases with two widely separated planets, the clustered periods model tends to find the higher peak in the relative likelihood associated with the inner planet (i.e., an inner cluster of planets has a higher probability of finding a planet close to the dynamical stability limit than an outer cluster of planets).  In systems with 3 or more planets, if the planets are grouped such that there are two clusters, then the overall likelihood of finding a planet tends to be higher nearer the multi-planet cluster than the single planet cluster.  If the planets are more regularly spaced and the clustered periods model deems them to be one wide cluster, then the likelihood of finding a planet tends to be evenly split across the cluster.  The planet radius likelihoods are the same across both period models.

\section{Discussion} \label{sec:discussion}

\subsection{Follow-up \textit{TESS} Observations}

A majority of the \textit{TESS} systems have their predicted orbital period maximum in relative likelihood at less than 40 days.  For many of these systems, the selection bias created by the \textit{TESS} observing pattern (roughly 27 days for each portion of the sky) produces a highly skewed distribution towards planets with shorter periods.  However, this also works to favor follow-up observations, as less observation time is required to cover the orbital period space.

The planet radius predictions are relatively flat across $1-3 R_\oplus$.  \textit{TESS} is sensitive to transits down to $\sim$ 100 ppm due to the benefits of being in space, but as it is magnitude-limited it is restricted to discovering transits in brighter (and usually therefore larger) stars, and correspondingly towards larger planets.  When this observation bias is coupled with the higher intrinsic likelihood of smaller planets across the relatively small sample here, the planet radius predictions become mostly flat with possibly a slight increase towards smaller planet radii.

\begin{figure}[ht]
    \centering
    \includegraphics[width=\columnwidth]{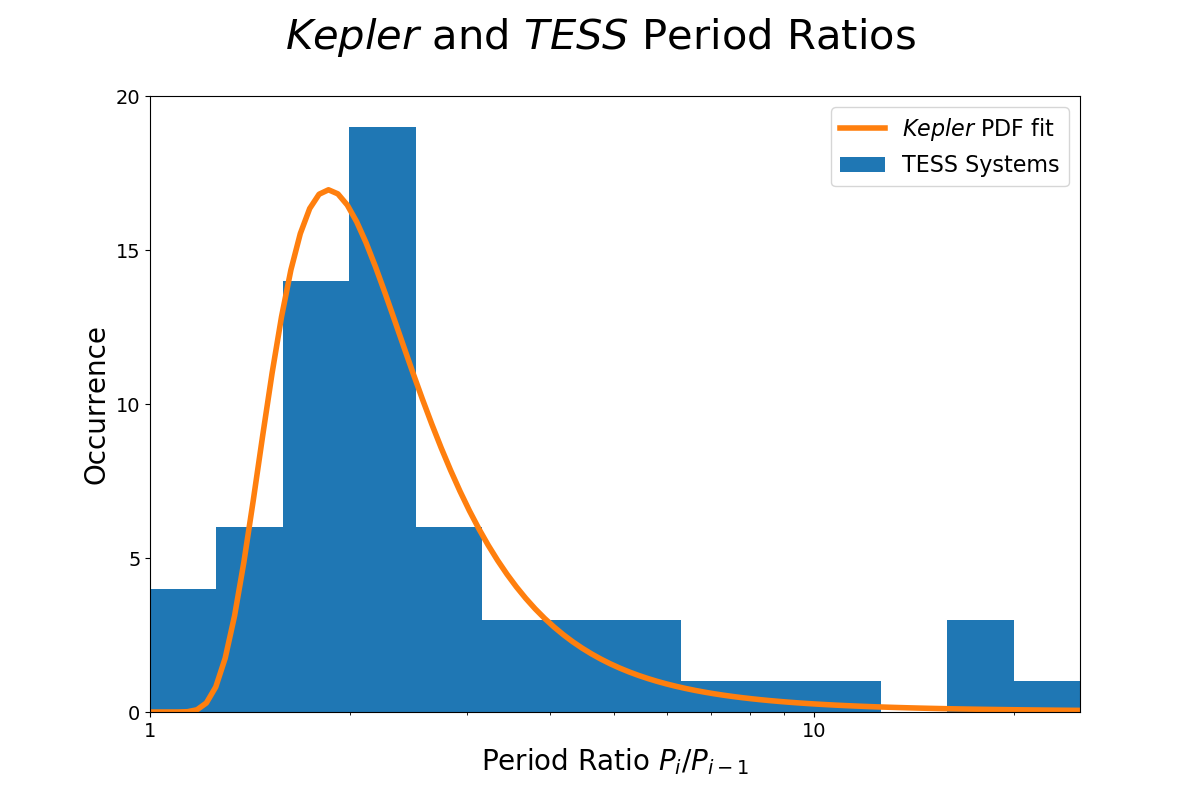}
    \caption{Period ratios from \textit{TESS} multi-planet systems, with the \textit{Kepler} period ratio distribution overlaid.  \textit{TESS} systems show an excess at $\sim 2.2$ compared to \textit{Kepler}, and the best fit lognormal distribution favors a slightly longer peak in the distribution.}
    \label{fig:TESS_Kepler_PRs}
\end{figure}

The mode of the adjacent-planet period ratios from the \textit{TESS} multi-planet systems in this sample is $\approx$ 2.2.  The \textit{TESS} data show an excess of period ratios there that are likely affected by the small sample size, as the rest of the data fits the \textit{Kepler} model (mode of planet ratios at $\sim 1.9$) fairly well given the sample size (see Figure \ref{fig:TESS_Kepler_PRs}).  For the \textit{Kepler} datasets, the period ratios between consecutive planets are fit by a lognormal distribution with parameters $\mu = -0.9$, $\sigma = 0.4$.  This parameter was previously described by a ``log10-normal distribution" from \citet{mal15} and \citetalias{mul18} with parameters $\mu = -0.39$, $\sigma = 0.18$; the shapes of these distributions are similar because our distribution simply scales the previous definition from \citetalias{mul18} to the standard lognormal distribution in the natural logarithm base.

\subsection{Period ratios vs clustered periods}

We tested two separate models for the period distribution in multi-planet systems, with the goal of using the predictions provided by each distribution to test which one is more accurate.  The period ratio model inserts planets according to the period ratios from \textit{Kepler} multi-planet system statistics, but taking into account only one insertion between planets and how that affects the period ratio between the injected planet and each of the known planets.  For example, for two planets with a period ratio of 7, this model finds that a planet injected at period ratio of $\sqrt{7} \approx 2.65$ has the highest probability, even though the \textit{Kepler} period ratios have a mode near 2.  This is because the probability of having two period ratios of 2.65 is more likely than the probability of having period ratios of 2 and 3.5.  This is still true for lower period ratio gaps; for example, a period ratio gap of $\sim$3 would find the probability of having two equal spaced gaps of $\sim$1.7 to be $0.93^2 \sim 0.86$, whereas the probability of having a gap of the period ratio mode $\sim$1.9 and a gap of $\sim$1.6 would be $1\times0.75 \sim 0.75$, relative to the mode having probability 1.  This model, therefore, tends to inject planets at even spacing between planets in log period space.

The clustered periods model inserts planets based on clusters in the orbital period where planets have been found.  However, due to the fact that the cluster locations in period space are fit from the currently known planet periods, the center of the cluster is likely to be on one planet.  Following the lognormal distribution, the probabilities will decrease outwards from the center of the cluster, so the period with the highest relative likelihood will be the period closest to the cluster center where the dynamical stability criterion is satisfied (likely exterior, as the lognormal distribution is right-skewed to longer periods).  This model prefers inserting planets as close as possible to known planets and tends to avoid the center of larger gaps between planets, as it views a gap as a cluster separator.

The two flavors of \mname{} (corresponding to the period ratios- or clustered periods-based predictions) can provide, for some systems, different predictions. Such differences reflect the inherent limitation of our understanding of the exoplanet population and, as our models are refined in the future, should disappear. In the interim, we provide here two thoughts on how such differences in predictions may be handled. In systems with highly regular spacing of planets with the exception of a single large gap, the period ratio model would likely give more accurate predictions. In contrast, in  systems with two planets clustered close and another one much further away, the clustered periods model might be preferable. For systems where a high-likelihood prediction is desired, the PDFs from the two models should be combined (multiplied) with equal weights. In contrast, the differences in predicted planets can be used to test the veracity of the period ratios vs. clustered periods description of the exoplanet population. For such tests, one should focus on systems where the greatest difference exists between the two predictions: testing the presence of planets in such systems will, then, provide the strongest evidence for which of the two descriptions approximate better the underlying planet population. Lastly, we point out that due to them modular implementation of \mname{}, the assumed period distribution can be readily updated as newer and more precise models are constructed, further improving \mname{}'s predictive power.

\subsection{Running model iteratively on one system}

The orbital period relative likelihoods can change significantly if \mname{} is run iteratively, which might be preferred in systems with large gaps (period ratio $\gtrsim 10$).  In this case, one planet is injected in this gap, and the injection method run for more than one iteration with the injected planet from the first run kept in the new system.  When this occurs, the positions near the known planets and the injected planet now have a higher relative likelihood.  For the period ratio model, this is because the period ratios are now much closer to the peak of the \textit{Kepler} distribution and therefore favor the existence of a planet.  For the clustered periods model, this is due to the first injected planet essentially collapsing a two-cluster model into a wide single-cluster model, which advocates for more planets in the middle of the cluster (in the remaining gaps between the known planets and the first injected planet).

In these certain cases where the system has a wide gap between planets in period space (i.e., TOI 703, TOI 713, TOI 1468), we find that injecting more than one planet in those gaps has a much higher relative likelihood than only one planet.  As we are normalizing all system injections to 1 planet, this spreads out the likelihood of finding exactly one planet inside these large period ratio gaps into a wide function that lowers the relative probability, making it seem as if there should be no injections in this gap.  If, however, \mname{} is run iteratively, we can recover multiple planets in these large gaps.  The smallest gaps in which \mname{} injects any planets are defined by the stability criterion. As long as the gap between the known planet and the injected planet is greater than 8 mutual Hill radii, \mname{} may inject a planet there; the relative likelihood is still defined by which period model is used.

\subsection{Robustness and sensitivity}

We tested the robustness and sensitivity of this method by running systems with known non-transiting planets or systems with known transiting planets removed.  The method was able to inject the known non-transiting planet or re-inject the removed planet with a high probability ($>95\%$) in the gap in period space between the two neighboring planets.  In most cases it was also able to place the planet with $>90\%$ likelihood in a smaller window around the correct period in all systems we used.

We also showed the predictive power of the method by analyzing the likelihood of possible or doubtful candidate signals being planets.  The method found that most candidates are likely planets based on the position of the signal in period space, even if other indicators show the candidate is unlikely.  This means the method will be sensitive to a large fraction of planets, although follow-up tests are necessary to determine the false alarm probability as many of these signals would likely be false positives.

\section{Summary} \label{sec:summary}

We presented a planet prediction model called \mname{} informed by population-level understanding of planetary systems, including orbital stability constraints and exoplanet population statistics and, in the future, predictions from planet formation models. \mname{} can predict the relative likelihood of the locations and size of yet-unknown planets. The key results of our study are as follows:

(1) We demonstrate that \mname{} can successfully predict the locations and sizes of known planets, when they are hidden from the algorithm.  The predictions are stable against perturbations in the planet periods that are an order of magnitude larger than the typical uncertainties.

(2) \mname{} can also assess the relative likelihood of the presence of one vs. several yet-unknown planets in a given system, as well as provide likelihood analysis of currently unconfirmed planet candidates. In such test cases \mname{} performed well.

(3) In general, we find that systems with gaps in the period ratio between consecutive planets larger than $\sim 3$ have high probability of inserting a planet in that gap.

(4) We apply \mname{} to the \textit{TESS} multi-planet systems and predict relative likelihoods for periods and sizes of yet-undetected planets that may exist in these systems. We identify those best-suited for follow-up observations (i.e., with shortest periods, highest transit probabilities, and deepest transits).

% A sentence or two of an ambitious vision for where this study may lead to:
\mname{} provides a modular framework for predicting the presence of yet-unknown planets. In addition to guiding follow-up transit and radial velocity observations now, \mname{} has the potential to -- with increasingly complete population-level understanding of planetary architectures -- guide future missions that aim to search for and characterize habitable zone Earth-like planets, whether via direct imaging like LUVOIR \citep{luv19} or HabEx \citep{gau20}, interferometry \citep[the LIFE mission;][]{qua19} or transits \citep[Nautilus Space Observatory,][]{apa19b}.

\acknowledgments

We acknowledge support from the Earths in Other Solar Systems Project (EOS), grant no. 3013511 sponsored by NASA.  This research has made use of the Exoplanet Follow-up Observation Program website, which is operated by the California Institute of Technology, under contract with the National Aeronautics and Space Administration under the Exoplanet Exploration Program.  We acknowledge use of the software packages NumPy \citep{van11} and SciPy \citep{vir19}.  This paper includes data collected by the Kepler mission and by the TESS mission.  Funding for the Kepler mission is provided by the NASA Science Mission Directorate, and funding for the TESS mission is provided by the NASA Explorer Program.  The citations in this paper have made use of NASA’s Astrophysics Data System Bibliographic Services. 

\bibliography{main}

\end{document}